\documentclass[pra,amsmath,amssymb,amsfonts,superscriptaddress,twocolumn]{revtex4-2}
\usepackage{graphicx,color,mathtools,bm,braket,newtxtext,newtxmath}

\providecommand{\abs}[1]{\left\lvert#1\right\rvert}

\usepackage[T1]{fontenc}
\usepackage{comment}
\usepackage{threeparttable}
\usepackage{booktabs}

\usepackage{ulem}

\usepackage[hyphenbreaks]{breakurl}
\usepackage[colorlinks=true,linkcolor=blue,citecolor=blue,urlcolor =blue]{hyperref}

\usepackage{appendix}
\usepackage{amsfonts,amssymb}
\usepackage{physics}
\usepackage{tikz}

\begin{document}
\title{Time-frequency Talbot effect as Clifford operations on entangled time-frequency GKP states}
\author{T. Pousset}
\affiliation{Télécom Paris-LTCI, Institut Polytechnique de Paris, 19 Place Marguerite Perey, 91120 Palaiseau, France}
\author{R. Dalidet}
\affiliation{Université Côte d’Azur, CNRS, Institut de physique de Nice, France}
\author{L. Labonté}
\affiliation{Université Côte d’Azur, CNRS, Institut de physique de Nice, France}
\author{N. Fabre}
\email{nicolas.fabre@telecom-paris.fr}
\affiliation{Télécom Paris-LTCI, Institut Polytechnique de Paris, 19 Place Marguerite Perey, 91120 Palaiseau, France}

\date{\today}

\begin{abstract}
The Talbot effect---a near-field diffraction phenomenon in which a periodic wavefront self-images at regular distances---can be transposed to the time--frequency domain via the space--time duality between diffraction and dispersive broadening.
We exploit this analogy to define the time--frequency (TF) Talbot effect and show that it implements different Clifford operations on TF Gottesman--Kitaev--Preskill (TF-GKP) qubits~(Phys.\ Rev.\ A~\textbf{102}, 012607), a class of qubit states encoded in the discretised frequency and time-of-arrival degrees of freedom of entangled photon pairs, whose logical basis corresponds to even and odd components of an entangled frequency comb.
These states are intrinsically robust against small frequency and temporal displacements, which can be further corrected by linear or nonlinear quantum error-correction schemes.
We analyse the role of the comb envelope and peak width relative to the free spectral range, and show that a compromise must be made between the gate fidelity of the Clifford gates induced by the TF-Talbot operation and the error-correction capacity of the code.
We then demonstrate that the signature of the TF-Talbot effect is directly accessible via the generalised Hong--Ou--Mandel interferometer: all six logical GKP states can be unambiguously distinguished by introducing a frequency shift of half the comb periodicity in one interferometer arm.
We conclude with a feasibility analysis based on current experimental technology, identifying the comb finesse as the key figure of merit for both gate performance and correctability. This conclusion extends naturally to quadrature GKP states, where a shear in quadrature phase space is precisely a Talbot effect.

\end{abstract}
\vskip2pc
\maketitle
\section{Introduction}

The Talbot effect is observed when a spatially coherent wave diffracts through a periodic aperture. This phenomenon occurs in the near-field diffraction regime, where identical and scaled replicas of the light distribution immediately after the aperture reappear at specific distances from it.
These distances are integer multiples of the Talbot length, defined as $z_{T} = 2l^{2}/\lambda$, where $l$ is the grating period and $\lambda$ is the wavelength of the incident wave.
For single-photon states, quantum Talbot carpets have been produced by directing twin photons through a periodic grating, demonstrating the possibility of performing logical operations on qudit states in the transverse position--momentum degree of freedom and defining a universal set of gates~\cite{barros_free-space_2017,farias_quantum_2015}.
Recent work has also explored the effect in structured light, such as optical vortices, where helical phase fronts give rise to intricate Talbot patterns with conserved orbital angular momentum~\cite{ikonnikov_two-dimensional_2020}.
In solid-state systems, tunnelling-induced interference in asymmetric quantum-dot molecules exhibits Talbot-like revivals, providing insight into coherent charge dynamics~\cite{azizi_tunneling-induced_2021}.
In the classical and quantum spatial domains, transverse position--momentum correlations of photon pairs or coherent fields replicate grating patterns at Talbot lengths~\cite{barros_free-space_2017}.
Temporal analogues exploit dispersion in optical fibres to mimic near-field diffraction: group-velocity dispersion imposes a quadratic spectral phase, enabling time--frequency self-imaging~\cite{Muriel:99,hall_temporal_2021}, which has been used for temporal cloaking, that is, the selective concealment of events through a time gap in a probe beam~\cite{lukens_temporal_2013}.
 \\

Time--frequency photon-pair comb states are produced by placing a spontaneous parametric down-conversion (SPDC) or four-wave mixing source inside an optical cavity~\cite{lu_mode-locked_2003,chang_648_2021,yoon_quantitative_2021,fabre_generation_2020,maltese_generation_2020,kues_quantum_2019}. Such frequency-entangled photon pairs are typically interpreted and exploited as entangled qudits.
An alternative description consists in interpreting such a state as a time--frequency GKP (TF-GKP) state, a fault-tolerant qubit code encoded in a frequency-comb entangled photon pair~\cite{fabre_generation_2020,maltese_generation_2020,fabre_teleportation-based_2023} (see also~\cite{PhysRevLett.130.200602,chang_gkp-inspired_2026}), and its generalisation to $n$-photon entangled states~\cite{descamps_gottesman-kitaev-preskill_2024}.
The TF-GKP state is designed to be robust against small frequency and temporal shifts, which can subsequently be corrected by linear or nonlinear error-correction schemes~\cite{fabre_teleportation-based_2023}.
The TF-GKP encoding is defined by analogy with the quadrature GKP state, which corresponds to the even and odd components of a coherent superposition of squeezed states of light~\cite{gottesman_encoding_2001,albert_performance_2018,fluhmann_encoding_2019,campagne-ibarcq_quantum_2020,larsen_integrated_2025,valahu_quantum-enhanced_2025}. In the TF-GKP case, the two logical codewords are defined by the coherent sum of even and odd frequency peaks, respectively.
Methods for performing logical operations within this frequency subspace have been proposed using time shifts~\cite{fabre_generation_2020} and pump engineering~\cite{maltese_generation_2020}.
Logical operations on two frequency bins can be implemented at the single-photon level~\cite{Lukens:17,lu_controlled-not_2019} or on photon pairs~\cite{PhysRevA.107.062610} with high fidelity ($>99\%$) by cascading electro-optic modulators and pulse shapers, or by exploiting the transverse degree of freedom with acousto-optic modulators~\cite{lukens_paradigm_2026}. However, these approaches come at the expense of reduced source brightness and do not extend straightforwardly to a larger number of frequency peaks.
A general proposal for performing high-fidelity, high-brightness logical operations on TF-GKP states therefore remains an open challenge.\\

In this paper, we show how the time--frequency Talbot effect, a frequency shear operation acting on entangled photon pairs, implements Clifford operations on TF-GKP qubits.
Although the Talbot effect has previously been proposed for logical operations on single-photon qudits in the transverse position--momentum~\cite{farias_quantum_2015,barros_free-space_2017} and time--frequency~\cite{Ogrodnik:25} degrees of freedom, its role as a logical gate on the even and odd comb components of a qubit encoding has not yet been identified.
A key point is that shears, rather than shifts, are required to remain within the physical TF-GKP subspace.
More specifically, the TF-Talbot effect implements the Clifford $\hat{X}_{t}$ operation on the frequency-entangled comb via a spectral chirp $\beta_{T} = \pi/\overline{\omega}^2$, where $\overline{\omega}$ is the comb periodicity. A chirp of $\beta_{T}/2$ instead induces the Clifford operation $\hat{S}\hat{R}_{y}(-\frac{\pi}{2})\hat{S}^{\dagger}$, where $\hat{S}$ is the phase gate and $\hat{R}_{y}(-\frac{\pi}{2})$ is a rotation by $-\pi/2$ about the $y$-axis of the Bloch sphere. Both gates require a single dispersive optical element: for a free spectral range of 40\,GHz, as for the source presented in \cite{chang_648_2021}, approximately 100\,km of optical fiber suffices, whether standard SMF-28 (0.2\,dB/km) or ultra-low-loss fiber, supplemented by a waveshaper to fine-tune the residual dispersion.

We consider both the envelope width and the peak linewidth of the comb relative to the FSR, corresponding respectively to timing and frequency noise, and derive the source requirements for a high gate fidelity ($\geq$95\%).
A gate fidelity below unity is not a limitation in this encoding, because the overlap errors can still remain correctable according to the Knill--Glancy framework~\cite{glancy_error_2006,fabre_teleportation-based_2023}, provided the time-frequency broadening, which acts as the noise in this encoding, stays below a threshold.
This built-in fault tolerance, absent from other photonic qubit encodings, is the distinctive advantage of the TF-GKP encoding, and is consistent with recent results on quadrature GKP states~\cite{brenner_composable_2025}.
We further point out that while infinitely narrow (in time) comb peaks would yield unit-fidelity operations in the case of infinitely many peaks, they are also more susceptible to dispersion, revealing that high-fidelity operation does not necessarily require narrow peaks (in time) when the number of peaks is finite. The signature of all six logical TF-GKP states is directly accessible via the generalised Hong--Ou--Mandel (HOM) interferometer. The coincidence probability directly samples the chronocyclic Wigner distribution~\cite{douce_direct_2013}, and the addition of a frequency shift of half the comb periodicity in one arm allows one to distinguish the four of the six logical states that cannot be resolved by temporal shifts alone.
We show in particular that the visibility of the dips and antidips present in the coincidence probability can be used to assess the fidelity of the operations.
We conclude with a review of experimental platforms suitable for generating TF-GKP states via cavity-enhanced type-II SPDC or via a spatial light modulator acting on broadband photon pairs, the latter offering a simpler near-term implementation. Crucially, TF-GKP states are substantially easier to generate than their quadrature counterparts~\cite{larsen_integrated_2025}, placing the present proposal within reach of many existing experimental groups.

The paper is organised as follows. We open in Sec.~\ref{TFform} with a review of the time--frequency single-photon formalism and the TF-GKP encoding, before establishing in Sec.~\ref{talbotmain} how the TF-Talbot effect acts on these states to implement logical operations. Sec.~\ref{HOM} then shows that all logical TF-GKP states generated by the TF-Talbot effect leave distinct signatures in the generalised HOM interferometer, provided a frequency shift of half the comb periodicity is applied in one arm. We turn in Sec.~\ref{Sectionexperimental} to a realistic experimental assessment of the proposal, identifying the comb finesse as the key figure of merit. Finally, Sec.~\ref{sec:conclusion} summarises our results and outlines perspectives.

\section{Time-frequency degrees of freedom of single photons}\label{TFform}
In this section, we define the grid states~\cite{fabre_generation_2020} upon which we will perform the logical operations, and we present the propagation operators for the logical gates.

\subsection{Time-frequency ideal GKP states}

We can describe a single photon with frequency $\omega$ in a given transverse mode $a$ with a bosonic creation operator, which acts on the vacuum state as $\hat{a}^\dagger_{a}(\omega)|\text{vac}\rangle = |\omega\rangle_{a}$. Furthermore, we define the annihilation operator through $\hat{a}_{a}(\omega)|\omega'\rangle_{a} = \delta(\omega - \omega')|\text{vac}\rangle$. Also, the commutation relation between the annihilation and creation operators is:
\begin{equation}
    [\hat{a}_{\alpha}(\omega),\hat{a}_{\gamma}^\dagger(\omega')] = \delta(\omega - \omega')\delta_{\alpha\gamma}\mathbb{I},
\end{equation}
where $\alpha$ and $\gamma$ designate other degrees of freedom, such as polarization. In what follows, we will omit this additional degree of freedom.
In the case where the central frequency of the spectral distribution is much larger than its spectral width~\cite{PhysRevA.72.032110,smith_photon_2007}, we can express the temporal annihilation operator as a function of frequency through the Fourier transform:
\begin{equation}
    \hat{a}(t) = \frac{1}{\sqrt{2\pi}}\int_{\mathbb{R}}\mathrm{d}\omega\hat{a}(\omega)\exp(-i\omega t).
\end{equation}
We now define the frequency operator as~\cite{fabre_time-frequency_2022}:
\begin{align}
    \hat{\omega}_{a}=\int_{\mathbb{R}} \mathrm{d}\omega \omega \hat{a}^{\dagger}(\omega) \hat{a}(\omega),
\end{align}
which has as eigenstate a single photon with frequency $\omega$: $\hat{\omega}_{a}\ket{\omega}_{a}=\omega\ket{\omega}_{a}$. This is a Hermitian operator, and therefore an observable. Even though negative frequencies are not observed, we have extended the integration range of the frequency operator to $\mathbb{R}$, since in practice, single-photon states will have a compact wavefunction far away from negative frequencies, and centered at GHz or THz frequencies. Then, the time-of-arrival operator is similarly defined:
\begin{equation}
    \hat{T}_{a}=\int_{\mathbb{R}} \mathrm{d}t t \hat{a}^{\dagger}(t) \hat{a}(t),
\end{equation}
which has as eigenstate a single photon with time-of-arrival $t$: $\hat{T}_{a}\ket{t}_{a}=t\ket{t}_{a}$, where $\ket{t}_{a}=\hat{a}^\dag(t)\ket{\mathrm{vac}}$. The time-of-arrival and frequency operators obey a Heisenberg algebra when the mode $a$ is occupied by a single photon~\cite{fabre_time-frequency_2022}:
\begin{equation}
[\hat{\omega}_{a},\hat{T}_{a}]=i\mathbb{I}_{a}.
\end{equation}
This algebra has important consequences for defining a universal set of gates using the time-frequency degrees of freedom of single photons~\cite{fabre_time-frequency_2022}, which is mathematically the same as for quadrature position-momentum variables~\cite{braunstein_quantum_2005}.\\

 A general wavefunction can be decomposed into the temporal or frequency basis, the eigenvectors of the time-of-arrival and frequency operators in the single-photon subspace, as:
\begin{equation}
    \ket{\psi}=\int_{\mathbb{R}} \mathrm{d}\omega \psi(\omega) \ket{\omega}_{a} = \int_{\mathbb{R}} \mathrm{d}t \tilde{\psi}(t) \ket{t}_{a},
\end{equation}
where $\tilde\psi(t) = \int_{\mathbb{R}}(\mathrm{d}\omega/\sqrt{2\pi})\psi(\omega)e^{-i\omega t}$ is the Fourier transform of $\psi(\omega)$.
\par The TF-GKP state is one way to discretize continuous variables for defining qubits. Let us discretize the frequency space into intervals of length $\overline{\omega}$, which will correspond to a free spectral range (FSR). In the spectral domain, the ideal TF-GKP states can be written as a coherent sum of Dirac distributions centered at even and odd multiples of $\overline{\omega}$~\cite{fabre_generation_2020}:
\begin{equation}
    \begin{aligned}
    \ket{\overline{0}_\omega} &=\sum_{n\in\mathbb{Z}} \ket{2n\overline{\omega}}_{a},\\
    \ket{\overline{1}_\omega} &=\sum_{n\in\mathbb{Z}} \ket{(2n+1)\overline{\omega}}_{a}.
    \end{aligned}
\end{equation}
  The probability amplitude in the frequency domain is: $\braket{\omega}{\overline0_\omega}=\sum_{n\in\mathbb{Z}}\braket{\omega}{2n\overline{\omega}}_\omega=\sum_{n\in\mathbb{Z}} \delta(\omega-2n\overline{\omega})$. In the temporal domain, the ideal time-frequency GKP states are:
\begin{equation}
    \begin{aligned}
        \ket{\overline{0}_\omega} &= \sum_{n\in\mathbb{Z}} \ket{\frac{\pi}{\overline{\omega}} n}_t=\ket{\overline{+}_t},\\
        \ket{\overline{1}_\omega} &= \sum_{n\in\mathbb{Z}} (-1)^n\ket{\frac{\pi}{\overline{\omega}} n}_t=\ket{\overline{-}_t},
    \end{aligned}
\end{equation}
where the Fourier transform naturally discretizes the temporal domain into intervals of length $\pi/\overline{\omega}$.
We point out that:
\begin{equation}
    \begin{aligned}
         \ket{\overline{+}_\omega}=\sum_{n\in\mathbb{Z}} \ket{n\overline{\omega}}_\omega= \sum_{n\in\mathbb{Z}} \ket{2n \frac{\pi}{\overline{\omega}}}_t=\ket{\overline{0}_t},\\
         \ket{\overline{-}_\omega}=\sum_{n\in\mathbb{Z}}  (-1)^{n} \ket{n\overline{\omega}}_\omega= \sum_{n\in\mathbb{Z}} \ket{(2n+1) \frac{\pi}{\overline{\omega}}}_t=\ket{\overline{1}_t}.\label{eq:minus_state}
     \end{aligned}
\end{equation}
 The $Y$-basis logical states, which arise naturally in the TF-Talbot evolution as we shall see, are defined as
\begin{equation}
    \begin{aligned}
        \ket{\overline{\pm i}_\omega} &= \ket{\overline0_\omega} \pm i\ket{\overline1_\omega} = \ket{\overline+_t} \pm i\ket{\overline-_t},\\
        \ket{\overline{\pm i}_t} &= \ket{\overline0_t} \pm i\ket{\overline1_t} = \ket{\overline+_\omega} \pm i\ket{\overline-_\omega}.
    \end{aligned}
\end{equation}
Ideal TF-GKP states are non-physical, as they consist of infinite sums of monochromatic (Dirac delta) states and are therefore not normalizable. It is precisely this non-normalizability that explains why the $1/\sqrt{2}$ prefactor does not appear in the equal superpositions above: for ideal GKP states, the Fourier
transform of $\ket{\overline{0}_\omega}$ produces a full time-domain comb $\sum_{n\in\mathbb{Z}}\ket{\tfrac{\pi n}{\overline{\omega}}}_t$, which already coincides with $\ket{\overline{+}_t}$ by definition, rather than as a superposition $\tfrac{1}{\sqrt{2}}(\ket{\overline{0}_t}+\ket{\overline{1}_t})$ with an explicit prefactor.
The factor $1/\sqrt{2}$ is a convention imposed at the qubit level only when one restricts to the finite-dimensional logical subspace spanned by normalizable TF-GKP states, where the inner products of the logical states are finite and superpositions must be properly normalized.
\par Finally, ideal TF-GKP states provide robustness against small time and frequency shift errors, similarly to their quadrature counterparts~\cite{gottesman_encoding_2001}.
Indeed, due to the spacing between the comb's peaks, the logical states remain orthogonal even after undergoing small time and frequency shifts.
This means that below a certain threshold on the amplitude of the shift, described later in the paper, the logical qubit space is protected from logical errors caused by this kind of physical error.

\subsection{Time-frequency physical GKP states}\label{sec:physical_gkp}
\begin{figure}
    \centering
    \includegraphics[width=0.45\textwidth]{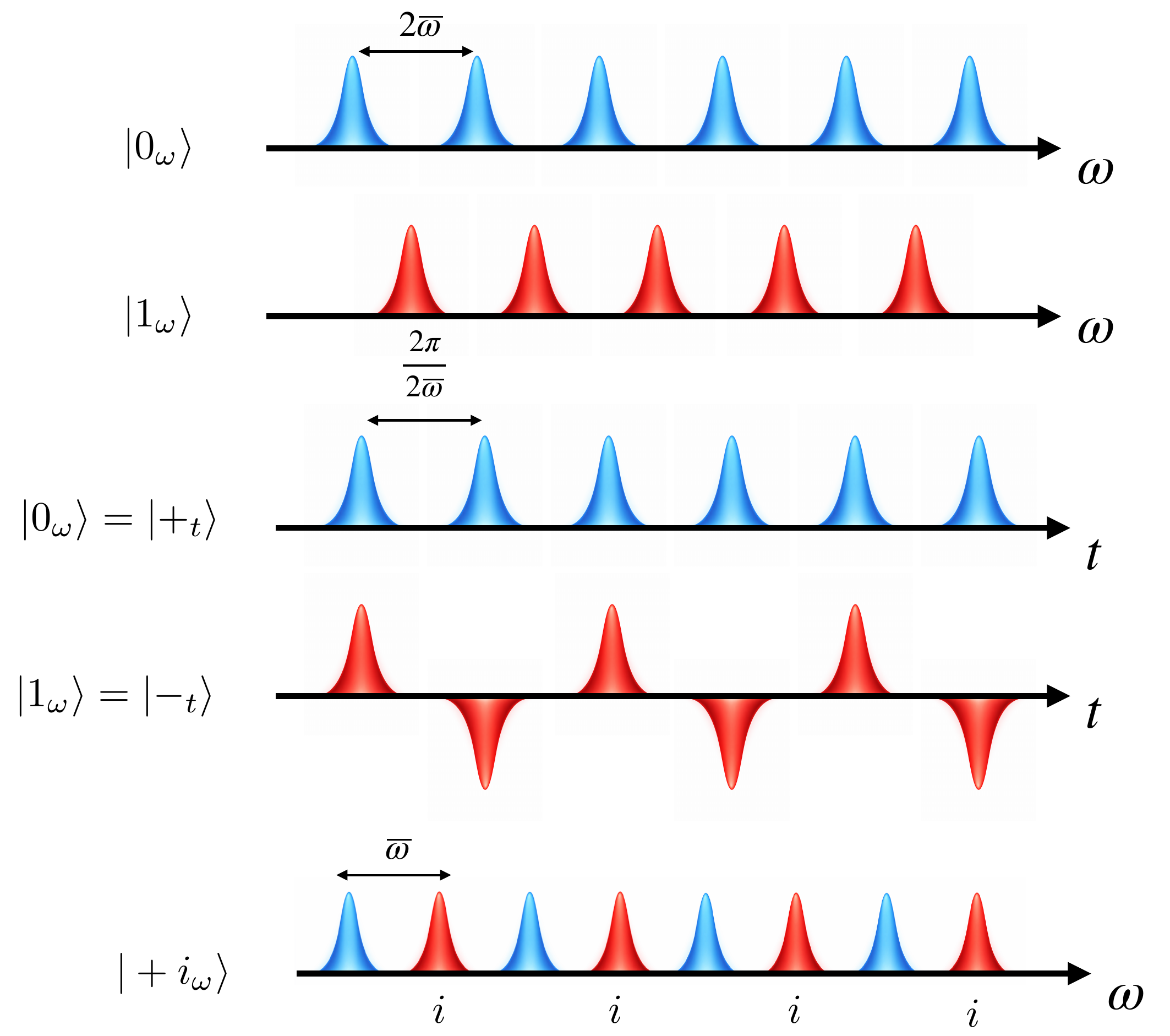}
    \caption{\label{wavefunctiongkp} Frequency-domain (top) and time-domain (bottom) amplitudes of the $\ket{0_{\omega}}=\ket{+_{t}}$ and $\ket{1_{\omega}}=\ket{-_{t}}$ logical codewords, together with the $\ket{+i_{\omega}}$ state, which arises naturally in the TF-Talbot evolution. The remaining logical states can be straightforwardly deduced from these.}
\end{figure}

Physical TF-GKP states can be obtained from ideal TF-GKP states by applying a Kraus-like operator~\cite{motes_encoding_2017,fabre_generation_2020}, and similarly for the other logical states:
\begin{equation}
    \ket{0_{\omega}}=\hat{K}\ket{\overline{0}_{\omega}}=\sum_{n\in\mathbb{Z}} \iint d\omega dt G_{\sigma}(\omega)G_{\kappa^{-1}}(t) \hat{D}(\omega) \hat{D}(t)\ket{2n\overline{\omega}}_{a},
\end{equation}
where $\hat{D}(\omega)$ and $\hat{D}(t)$ are displacement operators in frequency and time, respectively~\cite{fabre_generation_2020}, and the Gaussian envelope reads
\begin{equation}
G_{\sigma}(\omega)G_{\kappa^{-1}}(t)
=
\frac{1}{2\pi \sigma \kappa^{-1}}
\exp\!\left(
-\frac{\omega^2}{2\sigma^2}
-\frac{(t\kappa)^2}{2}
\right).
\end{equation}
Different logical states are summarized in Fig.~\ref{wavefunctiongkp}. The non-orthogonality of the logical states is an intrinsic property of a qubit defined as a discretization of continuous variables. The parameters $\sigma$ and $\kappa^{-1}$ play the role of frequency
and temporal noise widths, respectively, as made explicit when the auxiliary mode is populated by a single photon~\cite{fabre_time-frequency_2022}. In practice, $\sigma$ and $\kappa^{-1}$ receive contributions from distinct physical mechanisms. Temporal broadening arises from chromatic dispersion, polarization mode dispersion~\cite{antonelli_pulse_2005,poon_polarization_2008}, and temporal jitter~\cite{PhysRevLett.130.200602}, all of which increase $\kappa^{-1}$. Frequency broadening, which increases $\sigma$, originates from imperfect spectral manipulation, such as the finite resolution of waveshapers or the bandwidth limitations of electro-optic modulators.

Error correction of physical TF-GKP states is possible provided these noise parameters remain below the threshold defined in~\cite{glancy_error_2006} and, in our case, in Sec.~\ref{sec:knill-glancy}. Errors can be corrected using either linear (teleportation-based) or non-linear protocols~\cite{fabre_teleportation-based_2023}.

The $\ket{0_\omega}$ and $\ket{1_\omega}$ states can be written as follows~\cite{fabre_generation_2020}:
\begin{equation}
    \begin{aligned}
        \bra{\omega}\ket{0_\omega} &= N_{0_\omega}e^{-\frac{\omega^2}{2\kappa^2}}\sum_{n\in\mathbb{Z}}e^{-\frac{(\omega-2n\overline\omega)^2}{2\sigma^2}},\\
        \bra{\omega}\ket{1_\omega} &= N_{1_\omega}e^{-\frac{\omega^2}{2\kappa^2}}\sum_{n\in\mathbb{Z}}e^{-\frac{(\omega-(2n+1)\overline\omega)^2}{2\sigma^2}},\label{eq:physical_0_1}
    \end{aligned}
\end{equation}
where $N_{0_\omega}$ and $N_{1_\omega}$ are normalization factors whose values are detailed in Appendix~\ref{app:inner-products}.
In the large-envelope and small-peak-width limit, meaning $\kappa\gg\overline\omega\gg\sigma$, the normalization factors can be written as $N_{0_\omega}\approx N_{1_\omega}\approx\sqrt{2\overline\omega/(\pi\kappa\sigma)}$.
\par The overlap of the zero and one logical states is:
\begin{equation}
    \begin{aligned}
        &\braket{0_\omega}{1_\omega} \\
        &= N_{0_\omega}N_{1_\omega}\sqrt{\pi\frac{\kappa^2\sigma^2}{\kappa^2+\sigma^2}}\sum_{n,m\in\mathbb{Z}}e^{-\frac{(n-m-\frac{1}{2})^2\overline\omega^2}{\sigma^2}}e^{-\frac{(n+m+\frac{1}{2})^2\overline\omega^2}{\kappa^2+\sigma^2}},\\
        &\approx e^{-\frac{\overline\omega^2}{4\sigma^2}},
    \end{aligned}
\end{equation}
where the last line relies on the approximation that the envelope is large and the peak width is small compared to $\overline{\omega}$, meaning that $\kappa\gg\overline\omega\gg\sigma$ (see Appendix~\ref{app:inner-products} for detailed calculations).

\par Similarly, we can define the TF-GKP states in time as follows:
\begin{equation}
    \begin{aligned}
        \bra{t}\ket{0_t} &= N_{0_t}\sum_{k\in\mathbb{Z}} e^{-2\pi^2k^2\frac{\sigma^2}{\overline\omega^2}}e^{-\frac{(t-2k\frac{\pi}{\overline\omega})^2\kappa^2}{2}},\\
        \bra{t}\ket{1_t} &= N_{1_t}\sum_{k\in\mathbb{Z}} e^{-\frac{(2k+1)^2\pi^2}{2}\frac{\sigma^2}{\overline\omega^2}}e^{-\frac{(t-(2k+1)\frac{\pi}{\overline\omega})^2\kappa^2}{2}},\\
    \end{aligned}
\end{equation}
where the periodicity is $\pi/\overline\omega$, the peak width is $\kappa^{-1}$, the envelope width is $\sigma^{-1}$, and $N_{0_t}$ and $N_{1_t}$ are normalisation factors (see Appendix~\ref{app:inner-products} for details). In the large-envelope, narrow-peak limit $\kappa\gg\overline\omega\gg\sigma$, the normalisation factors reduce to $N_{0_t}\approx N_{1_t} \approx \sqrt{2\kappa\sigma/\overline\omega}$. By our convention of a continuous envelope in frequency space, the envelope of $\braket{t}{0_t}$ and $\braket{t}{1_t}$ is discrete in the time domain. Taking the Fourier transform, $\braket{t}{0_t}$ and $\braket{t}{1_t}$ map onto $\braket{\omega}{+_\omega}$ and $\braket{\omega}{-_\omega}$, defined as follows:
\begin{equation}
    \begin{aligned}
        \braket{\omega}{+_\omega} &= N_{+_\omega}e^{-\frac{\omega^2}{2\kappa^2}}\sum_{n\in\mathbb{Z}}e^{-\frac{(\omega-n\overline\omega)^2}{2\sigma^2}},\\
        \braket{\omega}{-_\omega} &= N_{-_\omega}e^{-\frac{\omega^2}{2\kappa^2}}\sum_{n\in\mathbb{Z}}(-1)^ne^{-\frac{(\omega-n\overline\omega)^2}{2\sigma^2}},
    \end{aligned}
\end{equation}
where $N_{+_\omega}$ and $N_{-_\omega}$ are normalization factors.
\par
\begin{figure}
    \centering
    \includegraphics[width=\linewidth]{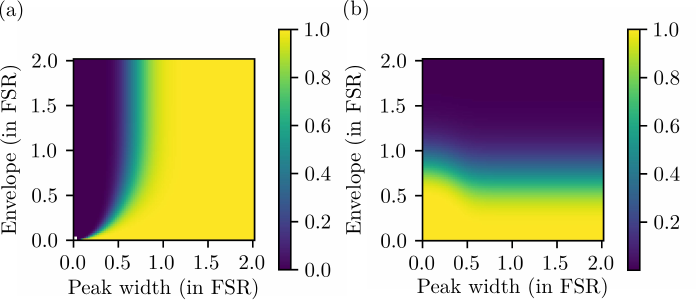}
    \caption{
  Overlap between the (a) $\ket{0_\omega}$ and $\ket{1_\omega}$ states and (b) $\ket{0_t}$ and $\ket{1_t}$ states. The physical TF-GKP states can be considered orthogonal only in the limit of a large envelope and narrow peaks.
    }
    \label{fig:overlap}
\end{figure}

 The overlap of the zero and one logical states in time is:
\begin{equation}
    \begin{aligned}
        &\braket{0_t}{1_t}\\
        &=N_{0_t}N_{1_t}\sqrt{\frac{\pi}{\kappa^2}}\sum_{k,l\in\mathbb{Z}}e^{-\frac{\pi^2\sigma^2}{2\overline\omega^2}((2k)^2+(2l+1)^2)}e^{-(k-l-\frac{1}{2})^2\frac{\pi^2}{\overline\omega^2}\kappa^2},\\
        &\approx e^{-\frac{\pi^2}{4}\frac{\kappa^2}{\overline\omega^2}},
    \end{aligned}
\end{equation}
where the last line relies on the approximation that the envelope is large and the peak width is small compared to the FSR, \textit{i.e}.\ $\kappa\gg\overline\omega\gg\sigma$ (see Appendix~\ref{app:inner-products} for details). In this limit, $\braket{0_\omega}{1_\omega}\approx 0$ and $\braket{0_t}{1_t}\approx 0$, recovering the ideal orthogonality. The exact overlaps $\braket{0_\omega}{1_\omega}$ and $\braket{0_t}{1_t}$ are shown in Fig.~\ref{fig:overlap}(a) and (b) for several values of the envelope width and peak width. The two plots are asymmetric, which we attribute to the envelope being continuous in frequency space but discrete in the time domain; the two representations become equivalent in the large-envelope, narrow-peak limit. In both cases, the overlap quantifies the probability of misidentifying the logical codewords and therefore serves as a figure of merit for the robustness of the qubit subspace against broadening-induced errors. The robustness of TF-GKP states against temporal and spectral displacements is analysed in Sec.~\ref{sec:knill-glancy}.

\subsection{Logical operations for TF-GKP states}

The single-qubit logical Clifford gates for ideal TF-GKP states can be cast as:
\begin{align}
\hat{X}_{\omega}=\hat{Z}_{t}=\ket{\overline{0}_\omega}\bra{\overline{1}_\omega}+\ket{\overline{1}_\omega}\bra{\overline{0}_\omega}, \\
\ \hat{Z}_{\omega}=\hat{X}_{t}=\ket{\overline{0}_\omega}\bra{\overline{0}_\omega}-\ket{\overline{1}_\omega}\bra{\overline{1}_\omega},
\end{align}
where we have observed that, since the Fourier transform of a comb is a comb, there is a redundancy of the logical operations in the temporal and frequency domains. The $\hat{Y}_{\omega}$ operation is defined as $\hat{Y}_{\omega}=i\hat{X}_{\omega}\hat{Z}_{\omega}$ (and similarly in the temporal domain).
For physical GKP states, the definition of the Pauli matrices is not that straightforward, due to phase issues that prevent the Pauli algebra from being satisfied. However, this can be tackled by considering protocols with a specific class of modular observables~\cite{asadian_heisenberg-weyl_2016,ketterer_quantum_2016}.

The TF-GKP encoding represents a discretization of frequency and time as continuous variables, and is therefore neither a discrete nor a continuous encoding~\cite{silva_periodic_2022}.
Consequently, the appropriate universal gate set for quantum computation is that of discrete qubits, rather than the native continuous-variable gates.
This distinction is justified~\cite{hastrup_cubic_2021} because a non-Gaussian resource such as the cubic phase gate contributes negligibly to the overall resource overhead. Moreover, applying such a gate to a GKP state typically maps it out of the protected code subspace. Therefore, universal fault-tolerant computation within the TF-GKP encoding employs a standard qubit-based gate set. This universal set comprises Clifford gates—such as the Hadamard and CNOT gates—supplemented by a non-Clifford gate, for instance, the $\pi/4$ gate.

The TF-Talbot effect acts as a shear operation that implements Clifford operations, as we shall see.

\subsection{TF-GKP entangled states}\label{TFentangledstate}

The generation of TF-GKP entangled states via an SPDC process in an optical cavity was proposed in Ref.~\cite{fabre_generation_2020}. We briefly recall the derivation here.
The wavefunction of a photon pair produced by a non-linear crystal from a type-II SPDC process is
\begin{equation}
    \ket{\psi}=\iint d\omega_{s} d\omega_{i} F(\omega_{s},\omega_{i}) \ket{\omega_{s},\omega_{i}}_{a,b},
\end{equation}
where $a,b$ are polarization modes, and $\omega_{s,i}$ are the signal and idler frequencies. The function $F$ is the joint spectral amplitude (JSA) and can be expressed as
\begin{equation}\label{eq:decomposition}
    F(\omega_{s},\omega_{i}) = f_{+}(\omega_{+})f_{-}(\omega_{-})f_{\text{cav}}(\omega_{s})f_{\text{cav}}(\omega_{i}),
\end{equation}
where $\omega_{\pm} = \frac{1}{\sqrt{2}}(\omega_{s}\pm\omega_{i})$ are called collective variables. Here, two functions are involved: $f_{+}$, which corresponds to the energy conservation condition and is derived from the pump beam's spectrum, and $f_{-}$, representing the phase-matching condition, which is determined by the spatial profile of the pump beam.
Additionally, the function $f_{\text{cav}}$ accounts for the cavity's effects and will be considered as the sum of Gaussian peaks:
\begin{equation}
     f_{\text{cav}}(\omega)=\sum_{n\in\mathbb{Z}} e^{-\frac{(\omega-n\overline{\omega})^{2}}{2\sigma^{2}}}.
\end{equation}
Note that the cavity transfer functions depend on the local variables $\omega_{s,i}$ rather than on the collective variables $\omega_{\pm}$, which affects the positions of the spectral peaks~\cite{fabre_teleportation-based_2023}.

The double Fourier transform of the joint spectral amplitude, denoted $\tilde{F}(t_{s},t_{i})$, is called the joint temporal amplitude, where $t_{s,i}$ are the times of arrival of the signal and idler photons. We will call $t_{\pm}=\frac{1}{\sqrt{2}}(t_{s}\pm t_{i})$ the temporal collective variables. \\

For continuous pumping, the function $f_{+}$ can be approximated by a Dirac distribution, $f_{+}(\omega_{+})=\delta(\omega_{+}-\omega_{p})$. As a consequence of this approximation, the JSA is a discretized line along the $\omega_{-}$ direction.
Let us consider the non-physical case where the frequency peaks are described by a Dirac comb: $f_{\text{cav}}(\omega)=\sum_{n\in\mathbb{Z}} \delta(\omega-n\overline{\omega})$.
By assuming that $f_{-}(\frac{(n-m)\overline{\omega}}{\sqrt{2}})=\exp(-(n-m)^2\overline{\omega}^{2}/(2\sigma^{2}))$ and $\overline{\omega}\gg \sigma$, we get $f_{-}(\frac{(n-m)\overline{\omega}}{\sqrt{2}})=\delta_{nm}$.
We point out that, due to continuous pumping, among the full grid, there are only even-even and odd-odd peaks along the $\omega_{-}$ axis, which is a TF-GKP entangled state. The wavefunction can therefore be written as:
\begin{equation}\label{TFGKPcollective}
    \ket{\psi}=\frac{\ket{R0_{\omega}0_{\omega}}_{ab}+\ket{R1_{\omega}1_{\omega}}_{ab}}{\sqrt{2}}=\sum_{n\in\mathbb{Z}} \ket{\omega_{-}=n\overline{\omega}},
\end{equation}
where $R$ denotes the restriction of the odd and even components to the antidiagonal only. In this last equation, we omit the frequency related to the energy conservation, which is modeled as a Dirac distribution as mentioned before.

\begin{figure}
    \centering
 \includegraphics[width=0.45\textwidth]{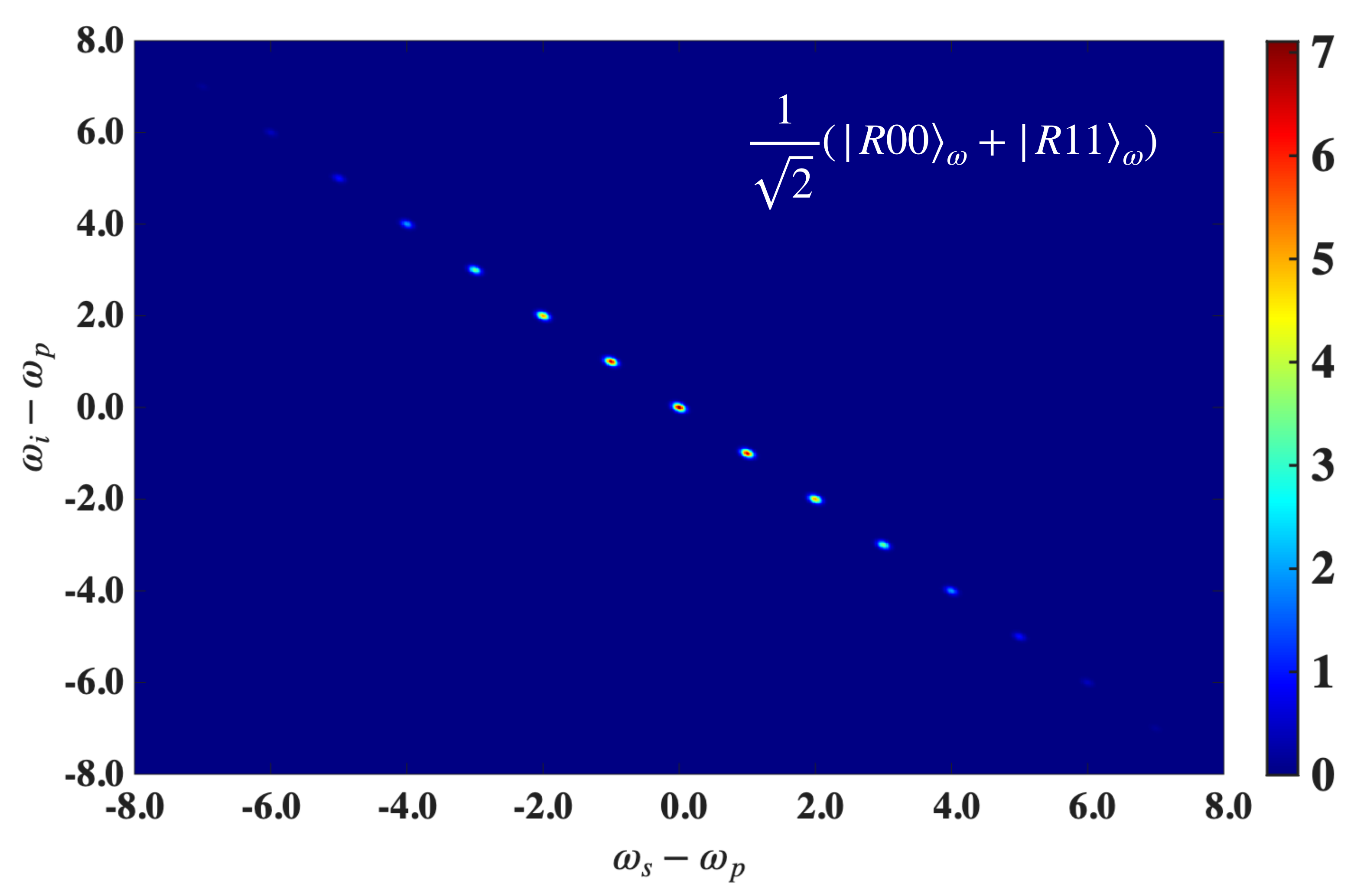}
\caption{\label{wavefunctiongkp1}Joint spectral intensity of the entangled TF-GKP state $\frac{\ket{R0_{\omega}0_{\omega}}_{ab}+\ket{R1_{\omega}1_{\omega}}_{ab}}{\sqrt{2}}$, with comb periodicity $\overline{\omega}=1$; $\omega_{p}$ is the pump degeneracy. The frequency units are set with respect to the free spectral range.}
\end{figure}

Following the same construction, the wavefunctions of physical TF-GKP entangled states are Gaussian rather than Dirac distributions.
An example of the resulting joint spectral intensity is shown in Fig.~\ref{wavefunctiongkp1}.
Although in this example each peak is narrow relative to the FSR, the small number of peaks---here ten, just for the sake of illustration---causes their temporal counterparts to overlap significantly, resulting in large timing noise; equivalently, $\kappa$ is not sufficiently large compared to $\overline{\omega}$.

\section{Time-frequency Talbot effect on photon pairs}\label{talbotmain}

\subsection{Propagation in the time-frequency domain}

In both the temporal and frequency domains, there is no natural analogue of diffraction. To emulate this effect, we induce diffraction-like behavior by applying a unitary transformation of the form $e^{i\beta \hat{\omega}^{2}}$.
The parameter $\beta$ plays the role of the propagation distance.
In essence, this operator imposes a quadratic phase modulation in the frequency domain—akin to the quadratic phase acquired by a spatial beam undergoing Fresnel diffraction.
The propagation of a single photon in a dispersive medium can be described by the following equation~\cite{agrawal_2022}:
\begin{equation}\label{equationdisp}
    (\frac{\partial^{2}}{\partial t^{2}}-i\frac{\partial}{\partial \beta}) \psi(\beta,t)=0 \ , \ -i\frac{\partial}{\partial \beta} \ket{\psi(\beta)} =\hat{\omega}^{2}\ket{\psi(\beta)},
\end{equation}
whose solution is:
\begin{equation}
    \ket{\psi(\beta)}=e^{i\beta \hat{\omega}^{2}} \ket{\psi(\beta=0)}.
\end{equation}
This transformation can be interpreted as a shear operation in the time-frequency chronocyclic phase space~\cite{fabre_generation_2020}. In direct analogy with the spatial domain, where a quadratic phase induces a shear in position-momentum phase space, here it produces a shear of the temporal and spectral profiles. The shear parameter is given by $\beta = k^{(2)}L$, where $L$ is the length of the dispersive medium and $k^{(2)}$ is its group-velocity dispersion (GVD). In what follows, we investigate the action of this propagator on entangled TF-GKP states.

\subsection{Time-frequency Talbot effect as a frequency shear}

We consider a TF-GKP entangled state in which each peak is infinitely narrow but has a finite envelope, which can be written as:
\begin{equation}
    \ket{\psi}=\ket{+_{\omega}}=\sum_{n\in\mathbb{Z}} c_{n}\ket{n\overline{\omega}}_{\omega_-},
\end{equation}
where the state is expressed in the collective variable $\omega_{-}$ as in Eq.~(\ref{TFGKPcollective}). Since we are in the continuous-pumping regime (see Sec.~\ref{TFentangledstate}), applying a frequency shear to one photon amounts to acting directly on the collective variable $\omega_{-}$ of the two photons. Therefore, after the propagation of Eq.~(\ref{equationdisp}), the wavefunction becomes:
\begin{equation}
e^{i\beta\hat{\omega}^{2}}\ket{\psi}=\sum_{n\in\mathbb{Z}} c_{n}e^{i\beta(n\overline{\omega})^{2}}\ket{n\overline{\omega}}_{\omega_-}.
\end{equation}
In the temporal domain, we obtain:
\begin{equation}
 \psi(t_{-},\beta) = \bra{t_{-}}\ket{\psi}= \exp(-t_{-}^{2}\kappa^{2}/2) \sum_{n\in \mathbb{Z}}e^{i\beta (n\overline{\omega})^{2}} e^{in\overline{\omega}t_{-}}.
\end{equation}
This is the cut of the joint temporal amplitude along the collective variable $t_{-}$ for different values of the frequency shear.
In the last equation, we have assumed that the envelope in the temporal domain is $\exp(-t_{-}^{2}\kappa^{2}/2)$, where $\kappa$ is the inverse of the temporal width.
This form of the envelope function coincides with the discretized one for infinitely large envelopes.
We can cast the previous equation as:
\begin{equation}\label{temporalwavefunction}
    \psi_{\ket{0_t}}(t_{-},\beta) = \exp(-t_{-}^{2}\kappa^{2}/2)  \sum_{n\in\mathbb{Z}}e^{in\overline{\omega}t_{-}}\exp(i\pi n^2 \beta/\beta_{T}),
\end{equation}
with
\begin{equation}
 \beta_{T} = \frac{\pi}{\overline{\omega}^2}.
\end{equation}

\subsubsection{One and two Talbot lengths as the $\hat{X}_{t}$ and $\mathbb{I}$ gates}\label{sec:ideal_X}

In this section, we demonstrate analytically that the TF-Talbot effect implements the $\hat{X}_t$ gate, equivalently the $\hat{Z}_{\omega}$ gate, on ideal TF-GKP states, and assess its performance numerically for physical TF-GKP states.

We begin with the ideal case. The wavefunction of Eq.~(\ref{temporalwavefunction}) expressed in the frequency domain reads:
\begin{equation}
    \psi_{\ket{\overline0_t}}(\omega_{-},\beta) = \sum_{n\in\mathbb{Z}} \delta(\omega_{-}-n\overline\omega)\,e^{i\pi\frac{\beta}{\beta_T}n^2}.
\end{equation}
After propagation through a dispersive element of chirp $\beta=\beta_T$, the odd frequency peaks acquire a phase of $\pi$, since $n^2$ and $n$ share the same parity. The wavefunction then becomes:
\begin{equation}
    \psi_{\ket{\overline0_t}}(\omega_{-},\beta_T) = \sum_{n\in\mathbb{Z}}(-1)^{n}\,\delta(\omega_{-}-n\overline\omega),
\end{equation}
which is precisely the $\ket{\overline{-}_{\omega}}$ state, or equivalently $\ket{\overline{1}_t}$, as defined in Eq.~(\ref{eq:minus_state}).\\

Secondly, since ideal TF-GKP states are $2\beta_T$-periodic under dispersion, applying a chirp $\beta = \beta_T$ to $\ket{\overline{1}_t}$ is equivalent to applying $\beta = 2\beta_T$ to $\ket{\overline{0}_t}$, which is precisely the identity: the state $\ket{\overline{1}_t}$ is thus mapped back to $\ket{\overline{0}_t}$. This periodicity is the time--frequency analogue of the Talbot effect for transverse position--momentum degrees of freedom~\cite{farias_quantum_2015,barros_free-space_2017}, where the field self-reproduces after propagating twice the Talbot length.

\subsubsection{Half Talbot length as the product of Clifford gates}\label{sec:ideal_H}

In this section, we calculate the phase change induced by the dispersion for $\beta=\beta_T/2$ and demonstrate that the TF-Talbot effect induces a product of Clifford gates.

A dispersion of $\beta_T/2$ performs the following transformations on the ideal TF-GKP codespace, as demonstrated in Appendix~\ref{appendix: halftalbot}:
\begin{equation}\label{equation:halftalbot}
    \begin{aligned}
        \ket{\overline0_t} &\rightarrow e^{i\frac{\pi}{4}}\ket{\overline{-i}_t} = e^{i\frac{\pi}{4}}(\ket{\overline0_t}-i\ket{\overline1_t}),\\
        \ket{\overline1_t} &\rightarrow e^{-i\frac{\pi}{4}}\ket{\overline{+i}_t} = e^{-i\frac{\pi}{4}}(\ket{\overline0_t}+i\ket{\overline1_t}).\\
    \end{aligned}
\end{equation}
\par Thus, a shear of $\beta_T/2$ stabilizes the ideal codespace and acts as a logical operation.
The transformation can be written in matrix form as:
\begin{equation}
    e^{i\frac{\beta_T}{2}\hat{\omega}^2} = \frac{e^{i\frac{\pi}{4}}}{\sqrt{2}}\begin{pmatrix}
        1 & -i\\
        -i & 1
    \end{pmatrix}_{(\ket{\overline0_t},\ket{\overline1_t})},
\end{equation}
where the $1/\sqrt{2}$ factor (as in the previous equation) is for normalization purposes since the operation must be unitary, and $e^{i\frac{\pi}{4}}$ is a global phase on the ideal TF-GKP codespace.
Finally, we write the transformation as a product of Clifford gates up to a global amplitude and phase factor:

\begin{multline}
    e^{+i\frac{\beta_T}{2}\hat{\omega}^2}=\\
    e^{i\frac{\pi}{4}}
    \begin{pmatrix}
        e^{-i\frac{\pi}{4}} & 0 \\
        0 & e^{i\frac{\pi}{4}}
    \end{pmatrix}
    \begin{pmatrix}
        \frac{1}{\sqrt{2}} & \frac{1}{\sqrt{2}}\\
        -\frac{1}{\sqrt{2}} & \frac{1}{\sqrt{2}}
    \end{pmatrix}
    \begin{pmatrix}
        e^{i\frac{\pi}{4}} & 0\\
        0 & e^{-i\frac{\pi}{4}}
    \end{pmatrix} = \hat{S}\hat{R}_{y}(-\frac{\pi}{2}) \hat{S}^{\dagger},
\end{multline}

where $\hat{S}$ is the $S$-phase gate, while $\hat{R}_{y}(-\frac{\pi}{2})$ denotes the rotation about the $y$-axis of the Bloch sphere by an angle of $-\pi/2$.

\begin{figure}
    \centering
    \includegraphics[width=\linewidth]{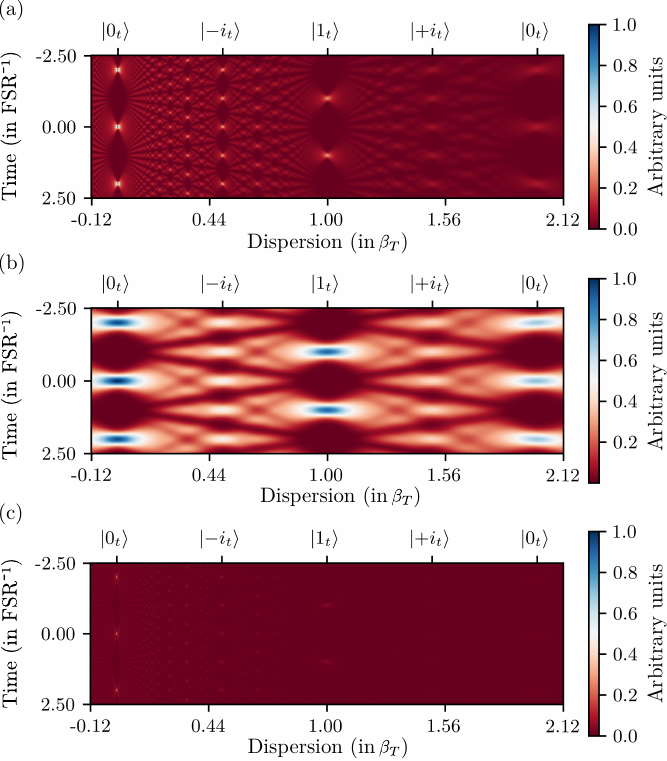}
    \caption{
    Joint temporal intensity along the $t_{-}$ collective variable as a function of dispersion, with (a) $(\kappa = 10\overline\omega, \sigma = 0.05\overline\omega)$, (b) $(\kappa = 2\overline\omega, \sigma= 0.05\overline\omega)$ and (c) $(\kappa = 30\overline\omega, \sigma = 0.05\overline\omega)$.
    The input state at $\beta = 0$ dispersion is a $\ket{0_t}$ state.
    The state is transformed into a $\ket{-i_t}$ state, a $\ket{1_t}$ state, then a $\ket{+i_t}$ state, and finally returns to being a $\ket{0_t}$ state.
    The smaller the envelope, the lesser the effect of dispersion for a fixed propagation.
    Hence, the transformation reaches higher fidelity for narrower envelopes.
    }
    \label{fig:talbot_realization}
\end{figure}

\subsubsection{Physical TF-GKP state and impact of dispersion}

The TF-Talbot effect for ideal and physical TF-GKP states is illustrated in Fig.~\ref{fig:talbot_realization}, where we plot the cut of the joint temporal intensity along the collective variable $t_{-}$, that is, $|\tilde\psi(t_{-},\beta)|^2$.
For ideal TF-GKP states, the shear operation shifts the temporal image within a fixed envelope, a qualitatively different action from a direct time delay, which translates the entire wavepacket (see~\cite{farias_quantum_2015} for an example in the transverse position-momentum domain).
For physical TF-GKP states, we study three scenarios sharing the same peak width $\sigma = 0.05\,\overline\omega$ but differing in their envelope: $\kappa = 10\,\overline\omega$ in Fig.~\ref{fig:talbot_realization}(a), $\kappa = 2\,\overline\omega$ in (b), and $\kappa = 30\, \overline\omega$ in (c).
In these cases, dispersion also temporally broadens the peaks, preventing the formation of perfect replicas.
The case of Fig.~\ref{fig:talbot_realization}(c) is particularly pathological.
The peaks are very narrow in time, which leads to a stronger broadening---for a fixed shear---compared with the two other cases, and the replicas are no longer visible after just one Talbot length.

\subsection{Gate fidelity of TF-Talbot effect and error robustness of physical TF-GKP states}

In this section, we present numerical results on the fidelity of the $\hat{X}_t$ and $\hat{S}\hat{R}_{y}(-\frac{\pi}{2}) \hat{S}^{\dagger}$ gates implemented via the TF-Talbot effect for the physical TF-GKP states introduced in Sec.~\ref{sec:physical_gkp}.

\subsubsection{GKP robustness against displacement errors}\label{sec:knill-glancy}

\begin{figure}
    \centering
    \includegraphics[width=\linewidth]{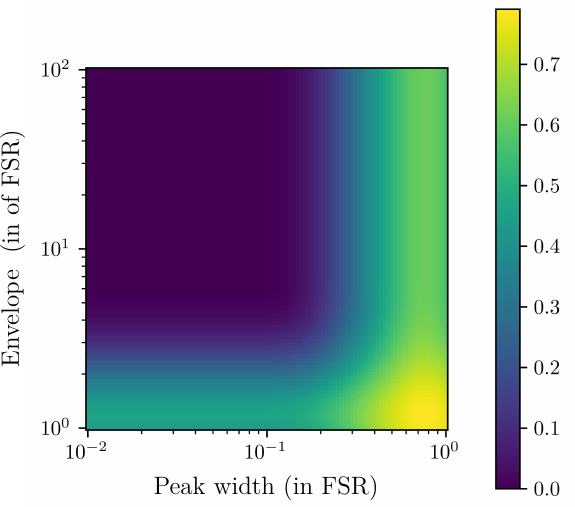}
    \caption{Error probability after Steane-type stabilisation. Probability that a noisy TF-GKP state undergoes a logical error after one stabilisation cycle using a physical TF-GKP ancilla in a Steane circuit. A higher error probability indicates reduced robustness of the TF-GKP state against displacement errors, and therefore a diminishing return from further stabilisation cycles.}
    \label{fig:knill-glancy}
\end{figure}

Robustness against displacement errors in the case of realistic GKP states is correlated with the overlap between Pauli states.
Indeed, the smaller the overlap, the larger the displacement needed to mix up a Pauli state with another.
However, this criterion is rather qualitative.
In this subsection, we consider the probability of having a logical error after a stabilization cycle of a TF-GKP state using a Steane-type correction circuit.
In the Steane-type error-correction circuit~\cite{gottesman_encoding_2001}, a noisy GKP state is corrected by entangling it with a fresher, less noisy ancilla GKP state via a CNOT gate. The ancilla is then measured with a quadrature-resolved (homodyne) detector, and a corrective displacement is applied to the data state conditioned on the measurement outcome. This corrects errors in one quadrature; a second, independent round, using a conjugate ancilla and the complementary homodyne measurement, then corrects the other quadrature. For GKP states encoded in the quadratures, it has been shown that for broadening smaller than $\sqrt{\pi}/6$, GKP states undergoing displacement errors can be stabilized using Steane-type error correction~\cite{glancy_error_2006,PhysRevA.93.012315}. Note that other error-correction schemes can be applied, such as~\cite{walshe_continuous-variable_2020,roy_decoding_2025}.
\par Thus, the probability of having an error after the stabilization process is the probability of having a broadening larger than $\sqrt{\pi}/6$.
For the TF-GKP case, the Steane and teleportation-based error-correction schemes have been described in~\cite{fabre_teleportation-based_2023}. The noise threshold in the TF-GKP encoding becomes $\overline\omega/6$ for frequency broadening and $\pi/(6\overline\omega)$ for time broadening.
This error probability is a figure of merit for the utility of subjecting a noisy GKP state to a stabilisation cycle using physical GKP ancillae: the larger the error probability, the less beneficial it is to stabilise the noisy state.

Using the modular variable formalism~\cite{ketterer_quantum_2016}, the probability of no error occurring after a stabilisation cycle can be expressed in terms of error functions~\cite{national_institute_of_standards_and_technology_nist_2010}:
\begin{multline}
    P_{\text{no error}}=\\
    \overline N_{0_\omega}^2\sum_{n,m,k,l\in\mathbb{Z}}e^{-\frac{\kappa^2}{4}(m-n)^2\frac{\pi^2}{\overline\omega^2}}e^{-\frac{(k-l)^2\overline\omega^2}{\sigma^2}}\\
    \times\sqrt{\frac{\pi}{4\kappa^2}}\left(\erf\!\left((3(m+n)+1)\frac{\pi\kappa}{6\overline\omega}\right)-\erf\!\left((3(m+n)-1)\frac{\pi\kappa}{6\overline\omega}\right)\right)\\
    \times\sqrt{\frac{\pi\sigma^2}{4}}\left(\erf\!\left((l+\tfrac{1}{6})\frac{\overline\omega}{\sigma}\right)-\erf\!\left((l-\tfrac{1}{6})\frac{\overline\omega}{\sigma}\right)\right),
\end{multline}
where $\erf(z) = \frac{2}{\sqrt{\pi}}\int_0^z\mathrm{d}x\,e^{-x^2}$ is the error function and $\overline{N}_{0_\omega}$ is a normalisation factor. Detailed derivations are provided in Appendix~\ref{app:knill-glancy}. In the regime $\kappa\gg\overline\omega\gg\sigma$, corresponding to a large envelope and narrow peaks relative to the FSR, the expression simplifies to:
\begin{equation}\label{probaerror}
    \mathrm{P}_{\text{no error}} = \erf\!\left(\frac{\pi\kappa}{6\overline\omega}\right)\erf\!\left(\frac{\overline\omega}{6\sigma}\right).
\end{equation}
We numerically evaluate the error probability $\mathrm{P}_{\text{error}} = 1 - \mathrm{P}_{\text{no error}}$ for several values of the envelope $\kappa$ and peak width $\sigma$ in Fig.~\ref{fig:knill-glancy}. As expected, the error probability decreases as the peaks become narrower and the envelope broader. This result stands in contrast with Fig.~\ref{fig:talbot_realization}, where a narrower envelope is seen to improve the fidelity of the Talbot operations $\ket{0_t}\rightarrow\ket{-i_t}$ and $\ket{0_t}\rightarrow\ket{1_t}$. We examine this trade-off in the following sections.

\subsubsection{State fidelity}

\begin{figure}
    \centering
    \includegraphics[width=\linewidth]{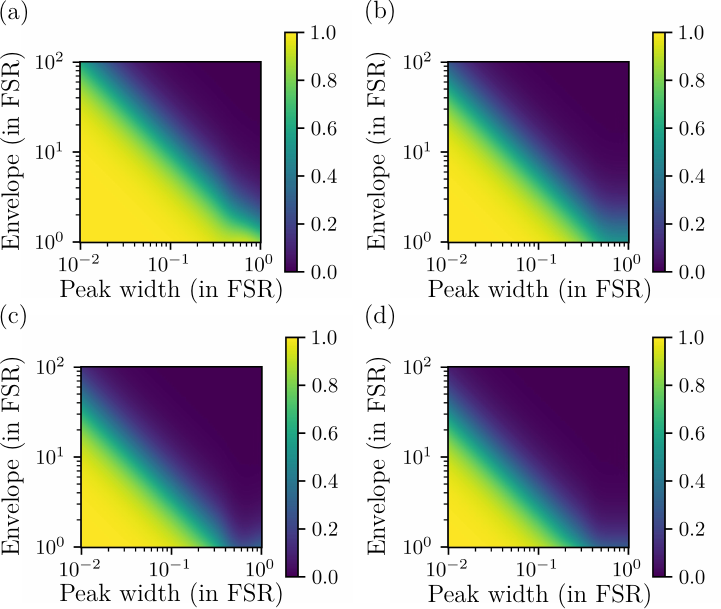}
    \caption{
    Fidelity of an initial $\ket{0_t}$ state undergoing the TF-Talbot effect with varying envelope and peak width \textit{vs}. (a) $\ket{-i_t}$ at dispersion $\beta=\beta_T/2$, (b) $\ket{1_t}$ at dispersion $\beta=\beta_T$, (c) $\ket{+i_t}$ at dispersion $\beta=3/2\times\beta_T$, (d) $\ket{0_t}$ at dispersion $\beta=2\beta_T$.
    The input state with peak width $\sigma$ and envelope $\kappa$ undergoes a shearing of $\beta$.
    We compute the fidelity of the traveling state with the TF-GKP states $\ket{+i_\omega}$, $\ket{1_t}$, $\ket{-i_\omega}$ and $\ket{0_t}$, all with peak width $\sigma$ and envelope $\kappa$.
    The fidelity is close to one in the case of narrow peaks and envelope.
    Indeed, while the TF-Talbot effect relies on interference between peaks of the grid state in the temporal domain, it is deteriorated by dispersion, which broadens the peaks in time.
    Dispersion is stronger on peaks with small duration, hence the fidelity increases when the envelope (in frequency) gets narrower.
    }
    \label{fig:fidelity_state}
\end{figure}

  We numerically simulate a physical TF-GKP state $\ket{0_t} = \ket{+_\omega}$ with envelope $\kappa$ and peak width $\sigma$ propagating through a dispersive element of increasing chirp $\beta$. Following the analytical results of Sec.~\ref{sec:ideal_X} and Sec.~\ref{sec:ideal_H}, we compute the fidelity between the evolved state and the four target codewords at each fractional Talbot length: $\ket{-i_t}$ at $\beta = \beta_T/2$ (Fig.~\ref{fig:fidelity_state}a), $\ket{1_t} = \ket{-_\omega}$ at $\beta = \beta_T$ (Fig.~\ref{fig:fidelity_state}b), $\ket{+i_t}$ at $\beta = 3\beta_T/2$ (Fig.~\ref{fig:fidelity_state}c), and $\ket{0_t}$ at $\beta = 2\beta_T$ (Fig.~\ref{fig:fidelity_state}d).

Two competing effects govern the fidelity as a function of the comb parameters. On the one hand, the TF-Talbot effect relies on constructive interference between multiple temporal pulses, so the larger the number of non-zero pulses, \textit{i.e}.\ the smaller $\sigma$ relative to $\kappa$, the higher the fidelity. On the other hand, each Gaussian pulse undergoes chromatic dispersion during propagation, leading to temporal broadening~\cite{agrawal_2022,saleh_2019}; narrower pulses broaden more strongly, so that for fixed $\sigma$, increasing $\kappa$ degrades the fidelity. As a result, for fixed $\sigma$ the fidelity decreases monotonically with $\kappa$, while for fixed $\kappa$ smaller values of $\sigma$ require a larger envelope to maintain the same fidelity threshold. Furthermore, since broadening accumulates with propagation distance, the fidelity decreases as $\beta$ increases: the constraints on $\kappa$ and $\sigma$ required to exceed a given fidelity threshold are therefore strictly more stringent at $\beta = 2\beta_T$ than at $\beta = \beta_T/2$.

These gate-fidelity constraints must be balanced against the requirements for error-correction robustness. As shown by the overlap analysis of Sec.~\ref{sec:physical_gkp} and the error probability computed in Sec.~\ref{sec:knill-glancy}, robustness against displacement errors demands a large envelope $\kappa$ and narrow peaks $\sigma$, which is precisely the regime in which gate fidelity is reduced. A compromise between the two must therefore be sought.

The three scenarios of Fig.~\ref{fig:talbot_realization} illustrate this trade-off concretely. In Fig.~\ref{fig:talbot_realization}(a), with $\kappa = 10\,\overline{\omega}$ and $\sigma = 0.05\,\overline{\omega}$, the JTI peaks remain well resolved and clearly visible after one Talbot length, yielding high gate fidelity while maintaining reasonable robustness against displacement errors; this represents the preferred operating point. Notably, although the state fidelity is strictly less than unity---as the state is not an ideal TF-GKP state---it remains within the correctable regime of the code. In Fig.~\ref{fig:talbot_realization}(b), reducing the envelope to $\kappa = 2\,\overline{\omega}$ raises the gate fidelity to $F \approx 0.9$ (Fig.~\ref{fig:fidelity_state}), but the state is no longer robust against temporal displacements, as evidenced by the non-zero error probability after Steane-circuit stabilisation in Fig.~\ref{fig:knill-glancy}. Conversely, in Fig.~\ref{fig:talbot_realization}(c), increasing the envelope to $\kappa = 30\,\overline{\omega}$ with the same peak width confers strong robustness against both temporal and spectral displacements, but the JTI peaks are no longer resolvable after one Talbot length, and the gate fidelity drops significantly (Fig.~\ref{fig:fidelity_state}).

\subsubsection{Gate fidelity}

\begin{figure}
    \centering
    \includegraphics[width = \linewidth]{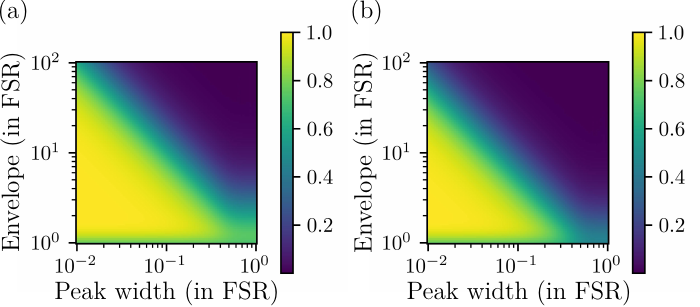}
    \caption{Gate fidelity against envelope and peak width used to define the physical GKP codespace for (a) the $\hat{S}\hat{R}_{y}(-\frac{\pi}{2}) \hat{S}^{\dagger}$ gate and (b) the $\hat{X}_t$ gate.}
    \label{fig:gate-fidelity}
\end{figure}

The study above only considers the fidelity given a fixed input state.
In order to consider all states, we use the gate fidelity metric~\cite{Lukens:17}.
For example, focusing on the $\hat{X}_t$ gate:
\begin{equation}
    \mathcal{F}_{\mathrm{gate}} = \frac{\mathrm{Tr}(\hat{W}^\dag\hat{X}_t)\mathrm{Tr}(\hat{X}_t^\dag\hat{W})}{\mathrm{Tr}(\hat{W}^\dag\hat{W})\mathrm{Tr}(\hat{X}_t^\dag\hat{X}_t)},
\end{equation}
where $\hat{W}$ designates the gate implementation and can be written as:
\begin{equation}
    \hat{W} = e^{i\beta\hat{\omega}^2}.
\end{equation}
\par We show the gate fidelity for both the $\hat{X}_t$ and $\hat{S}\hat{R}_{y}(-\frac{\pi}{2}) \hat{S}^{\dagger}$ gates, implemented with dispersions $\beta=\beta_T$ and $\beta = \beta_T/2$, in Fig.~\ref{fig:gate-fidelity}.
In both cases, the trace has been computed in the physical TF-GKP basis $\ket{0_t}$ and $\ket{1_t}$.
In order to account for the non-orthogonality of the $\ket{0_t}$ and $\ket{1_t}$ states, we first perform a Gram-Schmidt algorithm to construct an orthogonal basis and perform the trace using this orthogonal basis. High-fidelity operation ($\mathcal{F}_{\mathrm{gate}}>95$\%) can be obtained, as shown in Fig.~\ref{fig:gate-fidelity}.

\par As for the state fidelity computed in Fig.~\ref{fig:fidelity_state}, two competing effects govern the fidelity as a function of the comb parameters.
Since the TF-Talbot effect relies on interference between peaks in the temporal wavefunction, for a fixed envelope $\kappa$, a narrower peak width $\sigma$ yields higher fidelities, as it leads to more non-zero peaks in the temporal wavefunction.
Moreover, the TF-Talbot effect is deteriorated by dispersion, which broadens each peak in the temporal domain, hence the decrease for large envelopes (in frequency), as a large envelope means narrow temporal peaks, which are more subject to dispersive broadening.
Conversely, GKP robustness against temporal displacement errors requires large envelopes, so that the code can tolerate shifts without logical errors.
These two requirements are in tension: a compromise between gate fidelity and error-correction capacity must therefore be found when choosing the comb envelope.

\section{Obtaining the signature of the logical operation through the Hong--Ou--Mandel interferometer}\label{HOM}

\subsection{Signature of the TF-Talbot effect with the generalized HOM interferometer}\label{sec:HOM-theory}

\begin{figure}[h!]
\centering
\includegraphics[width=0.5\textwidth]{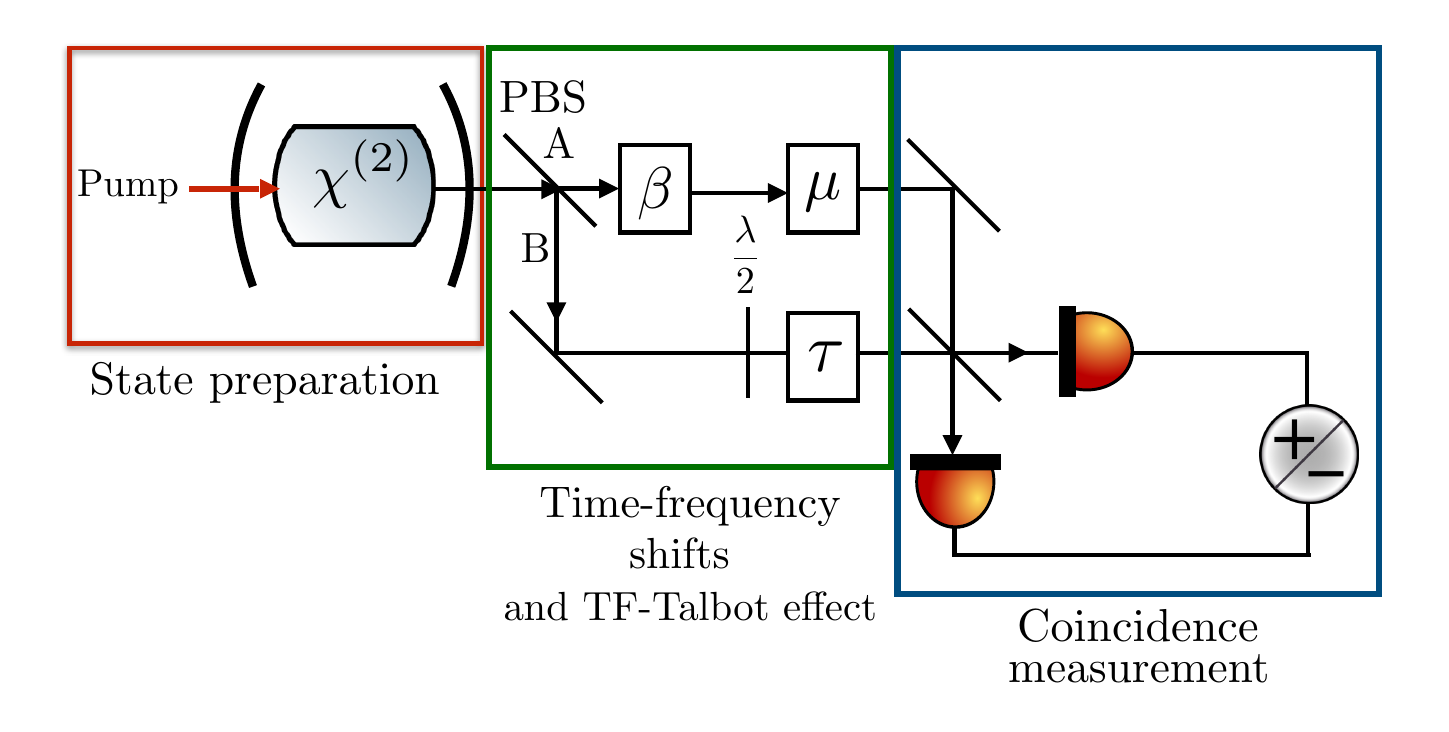}
\caption{\label{genHOM} Generalized HOM interferometer. After the preparation of a photon pair generated by a type-II SPDC process embedded in an optical cavity, the two photons are separated by a polarizing beam splitter (PBS). A shear operation $\beta$ is then applied to one of the photons, inducing the TF-Talbot effect. Each photon subsequently undergoes independent temporal $\tau$ and spectral $\mu$ shifts before being recombined on a balanced beam splitter. Finally, a time-unresolved coincidence measurement is performed.}
\end{figure}

\begin{figure*}[t]
    \centering
    \includegraphics[width=\linewidth]{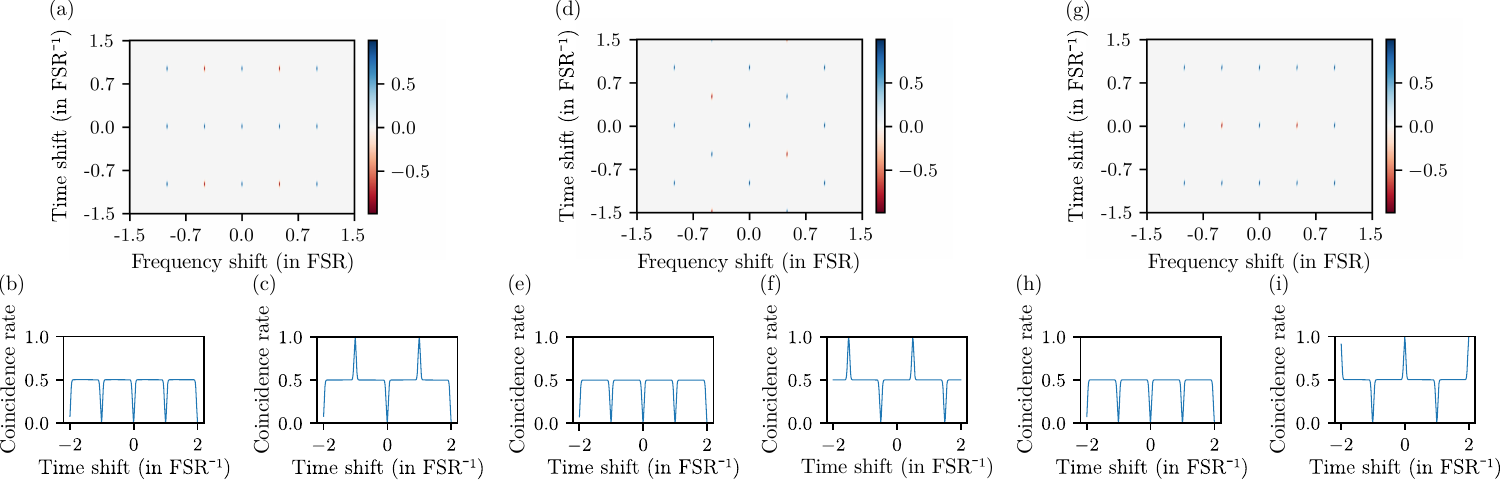}
    \caption{
    $\ket{0_t}$ state undergoing the TF-Talbot effect: (a) Wigner function at the origin, (b,c) HOM coincidence rate with no frequency shift and a half-FSR frequency shift, (d) Wigner function at half Talbot length, (e,f) HOM coincidence rate with no frequency shift and a half-FSR frequency shift, (g) Wigner function at one Talbot length, (h,i) HOM coincidence rate with no frequency shift and a half-FSR frequency shift.
    The state of interest is a physical $\ket{0_t}$ state of envelope $\kappa=10\overline\omega$ and peak width $\sigma=0.01\overline\omega$.
    The coincidence rates are computed without accounting for losses.
    The Wigner function at the origin (a) corresponds to a $\ket{0_t}$ state, the Wigner function at half Talbot length (d) corresponds to a $\ket{-i_t}$ state, and the Wigner function at one Talbot length (g) corresponds to a $\ket{1_t}$ state.
    Although the coincidence rate without frequency shift remains the same along the TF-Talbot effect, the one with a frequency shift of half an FSR presents varying patterns of dips and antidips.
    }
    \label{fig:wigner_talbot_origin}
\end{figure*}

We now show that the signature of the TF-Talbot effect can be observed with the generalized HOM interferometer.

The generalized HOM interferometer consists of applying independent temporal $\tau$ and frequency $\mu$ shifts to each photon produced by the SPDC process before recombining them on a balanced beam splitter followed by two single-photon detectors, as represented in Fig.~\ref{genHOM}.
The coincidence probability $I(\mu,\tau)$ is~\cite{douce_realistic_nodate}:
\begin{multline}
  I(\mu,\tau) = \tfrac{1}{2}\Biggl(1 -
    \Re\biggl[\iint d\omega_s\, d\omega_i \\
    \times \mathrm{JSA}^*(\omega_s{+}\mu,\omega_i)\,
    \mathrm{JSA}(\omega_s{+}\mu,\omega_i)\,
    e^{2i(\omega_s-\omega_i)\tau}\biggr]\Biggr).
\end{multline}
Assuming ideal energy conservation and using Eq.~(\ref{eq:decomposition}), this reduces to~\cite{douce_realistic_nodate,fabre_generation_2020}:
\begin{equation}
    I(\mu,\tau) = \frac{1}{2}\bigl(1-W_{-}(\mu,\tau)\bigr),
\end{equation}
where $W_{-}$ is the chronocyclic Wigner distribution of the state $\ket{\psi}$ of the photon pair along the collective frequency variable $\omega_{-}$, with wavefunction $F_-(\omega_-)$ written as follows:
\begin{equation}
    F_{-}(\omega_{-})=f_{-}(\omega_{-})\, f_{\mathrm{cav}}\!\left(\frac{\omega_{-}+\omega_{p}}{\sqrt{2}}\right)f_{\mathrm{cav}}\!\left(\frac{-\omega_{-}+\omega_{p}}{\sqrt{2}}\right).
\end{equation}
Finally, the chronocyclic Wigner distribution can be written as:
\begin{equation}
    W_{-}(\mu,\tau)=\int d\omega\, e^{2i\omega \tau} F_{-}(-\omega+\mu)F^{*}_{-}(\omega+\mu).
\end{equation}

Now, by inserting a dispersive optical element, such as an optical fiber or a waveshaper, into one arm of the interferometer, a quadratic spectral phase $e^{i\beta\omega_s^2}$ is applied to one photon.
Under the monochromatic-pump approximation, although the transformation is local, this shear maps the entangled TF-GKP codeword $\ket{0_t}$ onto the states $\ket{-i_t}$ and $\ket{1_t}$, and the signatures of all three states are directly accessible in the HOM coincidence pattern, as shown in Fig.~\ref{fig:wigner_talbot_origin}.

For the ideal TF-GKP states $\ket{\overline\pm_t}$ expressed directly in the collective variable $\omega_{-}$, the coincidence probability takes the form:
\begin{equation}
    I_{\ket{\overline\pm_t}}(\mu,\tau) = \frac{1}{2}\!\left(1 - \frac{1}{2}\sum_{(s,k)\in\mathbb{Z}^{2}}\delta\!\left(\tau-\frac{\pi s}{2\overline{\omega}}\right) \delta(\mu-\overline{\omega}k)\, e^{i\pi s(k+\nu)}\right),
\end{equation}
with $\nu=0$ for $\ket{\overline{+}_t}$ and $\nu=1$ for $\ket{\overline{-}_t}$. The corresponding expressions for the $\ket{\overline{0}_t}$, $\ket{\overline{1}_t}$ and $\ket{\overline{\pm i}_t}$ states are:
\begin{equation}
    \begin{aligned}
        I_{\ket{\overline0_t}}(\mu,\tau)&=\tfrac{1}{2}-\tfrac{1}{2}\sum_{s,k\in\mathbb{Z}}(-1)^{sk}
        \delta\big(\tau-\tfrac{\pi s}{\overline{\omega}}\big)\,\delta\big(\mu-\tfrac{k\overline{\omega}}{2}\big),\\
        I_{\ket{\overline1_t}}(\mu,\tau)&=\tfrac{1}{2}-\tfrac{1}{2}\sum_{s,k\in\mathbb{Z}}(-1)^{(s+1)k}
       \delta\big(\tau-\tfrac{\pi s}{\overline{\omega}}\big)\,\delta\big(\mu-\tfrac{k\overline{\omega}}{2}\big),\\
        I_{\ket{\overline{\pm i}_t}}(\mu,\tau)&=\tfrac{1}{2}-\tfrac{1}{4}\Big[\sum_{s,k\in\mathbb{Z}}\delta\big(\tau-\tfrac{\pi s}{\overline{\omega}}\big)\,\delta(\mu-k\overline{\omega})\\
        &\quad\pm\sum_{s,k\in\mathbb{Z}}(-1)^{k+s}\,\delta\big(\tau-\tfrac{(2s+1)\pi}{2\overline{\omega}}\big)
    \delta\big(\mu-\tfrac{(2k+1)\overline{\omega}}{2}\big)\Big],
    \end{aligned}
\end{equation}
see Appendix~\ref{app:wigner_ideal} for details.
The four ideal TF-GKP states, or the corresponding physical states $\ket{0_t}$, $\ket{1_t}$ and $\ket{\pm i}_t$, cannot be distinguished by temporal shifts alone; a frequency shift of $\overline{\omega}/2$ is required to distinguish them, as illustrated in Fig.~\ref{fig:wigner_talbot_origin}\,(a), (d), (g).
In particular, after applying such a frequency shift, the $\ket{0_t}$ and $\ket{1_t}$ logical states produce, respectively, a pattern of successive dips and a pattern of alternating dips and antidips.
Therefore, a full tomography of the chronocyclic Wigner distribution is not necessary to distinguish the six logical states~\cite{tischler_measurement_2015}.
When the coincidence probability exceeds $1/2$, the Wigner distribution is negative at that point, which constitutes a direct witness of entanglement~\cite{eckstein_broadband_2008}.
We note that a related HOM signature, obtained with a temporal shift (time delay) rather than a frequency shear, was reported in Ref.~\cite{fabre_generation_2020}.\\

\subsection{HOM visibility and gate fidelity}

\begin{figure}[t]
    \centering
    \includegraphics[width=\linewidth]{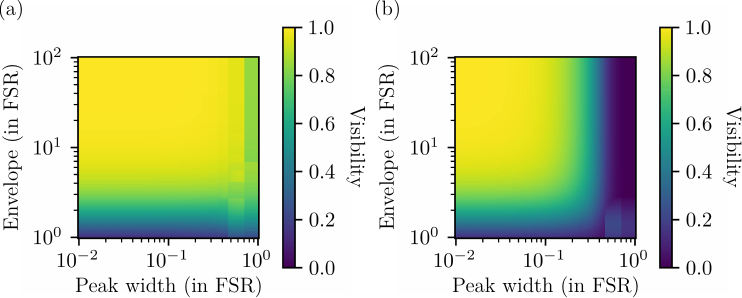}
    \caption{
    Visibility of the peak obtained with one positive frequency and time shift along the lattice of the chronocyclic Wigner function of the TF-GKP state after the TF-Talbot effect for (a) half a Talbot length and (b) one Talbot length.
    }
    \label{fig:visibility}
\end{figure}
To identify a Pauli TF-GKP state from HOM measurements, it suffices to resolve the coincidence rate over the elementary cell $(\tau,\mu)\in\bigl[-\frac{2\pi}{\overline\omega},\frac{2\pi}{\overline\omega}\bigr]\times\bigl[-\overline\omega,\overline\omega\bigr]$. Within this domain, the visibility of the dips and antidips in the coincidence probability provides a direct quantitative measure of the gate fidelity. Specifically, the visibility is related to the chronocyclic Wigner distribution by:
\begin{equation}
    V(\tau,\mu) = \frac{\left|\frac{1}{2}-I(\tau,\mu)\right|}{\frac{1}{2}} = \left|W_{-}(\tau,\mu)\right|.
\end{equation}

In Fig.~\ref{fig:visibility}, we plot the visibility of the peak adjacent to the origin obtained with one positive frequency and time shift along the lattice within the elementary cell for two target states: the $\ket{-i_t}$ state, obtained by propagating $\ket{0_t}$ through half a Talbot length (Fig.~\ref{fig:visibility}a), and the $\ket{1_t}$ state, obtained after one full Talbot length (Fig.~\ref{fig:visibility}b). Both are shown as a function of the envelope width $\kappa$ and peak linewidth $\sigma$, assuming a lossless setup. By overlaying Fig.~\ref{fig:gate-fidelity} and Fig.~\ref{fig:visibility}, one can simultaneously assess the gate fidelity from the HOM visibility and identify the parameter regime in which the state remains correctable, as characterised in Fig.~\ref{fig:knill-glancy}.

\section{Proposal for experimental implementation}\label{Sectionexperimental}

\subsection{Type-II cavity-based SPDC source}

We now survey representative experimental realizations of frequency-comb biphoton sources, with the aim of situating the parameter regimes---FSR, phase-matching bandwidth, and cavity linewidth---that are currently accessible.

Early demonstrations exploited integrated photonic platforms for their large and lithographically controllable FSRs.
A SiN microring resonator produced 40 frequency-mode pairs at a 50\,GHz FSR~\cite{imany_50-ghz-spaced_2018}, establishing the compatibility of CMOS-compatible waveguides with multimode entanglement.
In a similar spirit, a type-0 four-wave mixing Si-microring source achieved a mode spacing $\overline{\omega} = 2\pi\times 21$\,GHz with a cavity linewidth $\sigma = 2\pi\times 600$\,MHz, resolving up to 60 spectral peaks~\cite{PhysRevA.107.062610}. An AlGaAs type-II SPDC source with a monolithic built-in cavity reached $\overline{\omega} = 2\pi\times 19.2$\,GHz with a total phase-matching bandwidth $\kappa = 2\pi\times 10.9$\,THz and a cavity linewidth $\sigma \simeq 2\pi\times 0.02$\,THz~\cite{fabre_generation_2020}. A hybrid approach combining a PPLN waveguide with a fiber Fabry--P\'erot cavity demonstrated that the cavity finesse and the nonlinear conversion can be optimised independently, yielding a 15\,GHz FSR and a 0.5\,GHz linewidth~\cite{cheng_design_2019}.\\

The source best suited to observing the TF-Talbot effect, and the one that simultaneously satisfies all the requirements identified in the previous section, is that of Ref.~\cite{chang_648_2021}. This PPKTP type-II cavity source operates at 1316\,nm and combines a phase-matching bandwidth of $\kappa=$245\,GHz with an FSR of 45.32\,GHz and a cavity linewidth of $\sigma$= 1.56\,GHz, resolving 648 spectral peaks.
Crucially, these parameters fall in the regime $\kappa \gg 2\overline{\omega}$ with $\sigma = 0.05\,\overline{\omega}$, where the TF-Talbot effect implements Clifford operations with close-to-unit fidelity while the state remains correctable.

\begin{table*}[t]
\centering
\footnotesize
\begin{tabular}{@{}l cc c cccc cc cc@{}}
\toprule
 & & & $P_{\mathrm{error}}$
 & \multicolumn{4}{c}{Fidelity $F$}
 & \multicolumn{2}{c}{Gate fidelity $\mathcal{F}_{\mathrm{gate}}$}
 & \multicolumn{2}{c}{Visibility $\mathcal{V}$} \\
\cmidrule(lr){5-8} \cmidrule(lr){9-10} \cmidrule(lr){11-12}
Ref. & $\kappa$ & $\sigma$ &
 & $\beta_{T}/2$ & $\beta_{T}$ & $3\beta_{T}/2$ & $2\beta_{T}$
 & $\beta_{T}/2$ & $\beta_{T}$
 & $\beta_{T}/2$ & $\beta_{T}$ \\
\midrule
\cite{chang_648_2021} & $5.40$  & $0.03$  & $3.92\times10^{-3}$
 & $0.983$ & $0.934$ & $0.864$ & $0.781$
 & $0.983$ & $0.934$
 & $0.936$ & $0.932$ \\
\cite{fabre_generation_2020} & $567.7$ & $1.04$  & $0.550$
 & $1.89\times10^{-4}$ & $3.25\times10^{-5}$ & $1.43\times10^{-4}$ & $1.19\times10^{-4}$
 & $1.26\times10^{-4}$ & $3.25\times10^{-5}$
 & $0.915$ & $2.70\times10^{-5}$ \\
Sec.~\ref{sec:SLM} & $10$    & $0.32$  & $0.307$
 & $0.125$ & $0.0353$ & $0.155$ & $8.75\times10^{-3}$
 & $0.128$ & $0.0353$
 & $0.976$ & $0.556$ \\
\cite{PhysRevA.107.062610} & $60$    & $0.028$ & n.p.
 & $0.347$ & $0.117$ & $0.0556$ & $0.0321$
 & $0.347$ & $0.117$
 & $0.9994$ & $0.9954$ \\
\bottomrule
\end{tabular}
\caption{Simulated performance metrics for four representative operating
points $(\kappa,\sigma)$, given in units of the cavity FSR in full width at half maximum format: error probability $P_{\mathrm{error}}$ (Fig.~\ref{fig:knill-glancy}), state
fidelity $F$ (Fig.~\ref{fig:fidelity_state}), gate fidelity $\mathcal{F}_{\mathrm{gate}}$ (Fig.~\ref{fig:gate-fidelity}),
and interference visibility $\mathcal{V}$ (Fig.~\ref{fig:visibility}), evaluated at
fractions of the shearing $\beta_{T}$. For $\kappa=60$, $\sigma=0.028$,
$P_{\mathrm{error}}$ (denoted n.p.) lies below the numerical precision
of the simulation.}
\label{tab:performance_summary}
\end{table*}

\subsection{Type-II SPDC source followed by a spatial light modulator}\label{sec:SLM}
We now describe a promising alternative route to the generation of time--frequency GKP states based on a type-II SPDC source.

The type-II SPDC source is based on a periodically poled KTP (PPKTP) bulk crystal generating degenerate photon pairs at telecom wavelengths around $1544~\mathrm{nm}$ ($\simeq 194.2~\mathrm{THz}$), pumped by a picosecond mode-locked laser at $772~\mathrm{nm}$ with an average power of a few hundred milliwatts~\cite{bruno_generation_2014}. The crystal is $3~\mathrm{cm}$ long and the poling period ($\Lambda \simeq 47.8~\mu\mathrm{m}$) is chosen to fulfill the group-velocity-matching condition, enabling the generation of spectrally uncorrelated signal and idler photons. The spectral bandwidth can be increased up to about $10~\mathrm{nm}$ by reducing the crystal length. The optimal pump bandwidth is approximately $0.33~\mathrm{nm}$ (corresponding to a pulse duration of $\simeq 2.6~\mathrm{ps}$), which results in a JSI with an almost circular Gaussian shape and a spectral purity of $P \simeq 0.91$ without any spectral filtering. The photons are deterministically separated by polarization and efficiently coupled into single-mode telecom fibers with a coupling efficiency of about $90\%$ per photon. In this configuration, the high indistinguishability is confirmed by Hong--Ou--Mandel interference visibilities around $91\%$.

One convenient way to generate such grid states is to use a programmable spectral filter, as commonly employed in the community, such as a Coherent Waveshaper (4000B/C series). Although the maximum spectral resolution can reach the $1~\mathrm{GHz}$ level, it is often advantageous to relax this constraint in order to increase the transmitted coincidence-pair rate. In our case, the practical operating point of the spatial light modulator corresponds to spectral bins of about $330~\mathrm{pm}$ (\textit{i.e}.\ $\simeq 40~\mathrm{GHz}$) with an FSR of $1~\mathrm{nm}$ ($\simeq 125~\mathrm{GHz}$). This choice represents a compromise between moderate insertion losses (about $5~\mathrm{dB}$) and sufficiently large spectral bandwidths to maintain coincidence rates of a few kilohertz. Narrower spectral peaks would indeed improve the spectral finesse of the grid state, but at the cost of a drastic reduction of the coincidence rate, leading to prohibitively long acquisition times. We note that HOM interference at a 100\,GHz FSR has recently been demonstrated using single-sideband modulation to align comb-tooth frequencies at the single-photon level~\cite{yan_ten-channel_2025}, confirming the viability of this operating regime.

This source is, however, too noisy in the temporal domain, as we do not have $\kappa \gg \overline{\omega}$ for a GKP state robust against time-frequency shifts: the phase-matching bandwidth does not support enough spectral peaks.
In addition, the peak linewidth is too large ($\sigma\approx0.3\overline\omega$), so that not enough peaks are present in the temporal wavefunction for the Talbot interference effect to take place.
Increasing the bandwidth through phase-matching engineering, for instance by shortening the crystal or by adopting an aperiodically poled design, would be the natural path to overcome this limitation.

\subsection{Inducing dispersion with optical fiber and waveshapers}

We consider an operating point of 40\,GHz FSR, consistent with the source of Ref.~\cite{chang_648_2021}.
The frequency shear required for the Talbot effect can be implemented through several approaches.
A programmable waveshaper can directly imprint the required spectral phase with a resolution of 1\,GHz (10\,pm), at the cost of approximately 5\,dB of insertion loss; future integrated designs based on photonic foundry PDK data suggest this could be reduced to $\leq 1$\,dB~\cite{lu_controlled-not_2019,aim_pdk_2018}.
Alternatively, the quadratic spectral phase can be accumulated passively through propagation in a dispersive medium, such as dispersion-engineered optical fibers, chirped fiber Bragg gratings (e.g.\ commercially available from Proximion), or integrated photonic circuits with engineered group-velocity dispersion.
For a 40\,GHz FSR, this corresponds to approximately 100\,km of standard SMF-28 fiber, incurring 20\,dB of propagation loss; ultra-low-loss fiber (0.14\,dB/km) reduces this to 14\,dB, to which the insertion loss of a waveshaper used to fine-tune the residual dispersion must be added ($\approx 5$\,dB).
The total insertion loss of the frequency-shear stage therefore ranges from $14 + 5 = 19$\,dB (ultra-low-loss fiber) to $20 + 5 = 25$\,dB (standard SMF-28).
For comparison, implementing logical operations on frequency-bin qubits by concatenating two pulse shapers and one EOM incurs up to 13\,dB of insertion loss~\cite{Lukens:17}, with the additional requirement of complex numerical optimisation.
The present approach avoids this optimisation overhead, but currently entails higher passive losses.

\subsection{Performing the frequency shift}
Achieving a controlled frequency shift equal to half the FSR is a central experimental challenge. The accessible frequency range and the associated technical constraints differ depending on the target shift magnitude: shifts in the 0--25\,GHz range are routinely achievable, whereas the 25--100\,GHz regime remains significantly harder.

The thermo-optic effect of the nonlinear crystal can be exploited to shift the cavity resonances. This technique was employed in Ref.~\cite{tischler_measurement_2015}, where the full chronocyclic Wigner distribution was sampled. EOM-based solutions offer faster control but introduce approximately 2--3\,dB of insertion loss from input/output coupling alone. Driving an electro-optic phase modulator with a sawtooth waveform (serrodyne) produces a single-sideband frequency shift equal to the ramp repetition rate, at the cost of residual sidebands from the finite flyback time. This technique requires synchronisation between the driving electronics and the laser. A birefringence-based implementation was demonstrated in Ref.~\cite{kurzyna_variable_2022}, achieving a tuning coefficient of $\mu/V_\mathrm{pp} \approx 6$\,GHz/V (see also Ref.~\cite{golestani_electro-optic_2022}). Serrodyne operation spanning 200\,MHz to 1.2\,GHz was reported in Ref.~\cite{Johnson:10}.

Another solution is to employ an IQ modulator, \textit{i.e}., two electro-optic phase modulators embedded in a dual-parallel Mach--Zehnder interferometer driven by sinusoidal RF signals in phase quadrature, which can be biased in the carrier-suppressed single-sideband (CS-SSB) configuration to cancel one sideband and suppress the optical carrier. Chen~\textit{et al.}~\cite{chen_single-photon_2021} achieved a 15.65\,GHz blue shift with 15\,dB of insertion loss; a 3.4\,GHz shift was demonstrated in Ref.~\cite{ullah_digital_2024}.

\subsection{Single-photon detector requirements}

Another key parameter is the timing jitter of the superconducting nanowire single-photon detectors (SNSPDs).
The temporal separation between the interference features (or ``dips'') must exceed the detector jitter in order to be resolved.
Reducing the jitter generally comes at the expense of detection efficiency, so a trade-off must be found.
A realistic operating point is a timing jitter of about $25~\mathrm{ps}$ together with a system detection efficiency of around $80\%$, values that are representative of state-of-the-art commercial SNSPDs (e.g.\ from Single Quantum).
Such a jitter corresponds to resolvable temporal delays compatible with FSRs on the order of $40~\mathrm{GHz}$, and HOM dips were observed with the setup of~\cite{chang_648_2021}.
Finally, we note that state-of-the-art SNSPDs at telecom wavelengths have reached a timing jitter of 4.3\,ps~\cite{korzh_demonstration_2020}, below the 25\,ps period set by a 40\,GHz FSR, and would allow one to resolve dips and antidips for TF-GKP states of envelope narrower than $\kappa\lessapprox 250$\,GHz.

\section{Conclusion}\label{sec:conclusion}

We have shown that the TF-Talbot effect provides a framework for implementing Clifford logical gates on TF-GKP qubits encoded in entangled photon pairs. Different frequency shears on the collective variables of the photon pair induce the Clifford operations $\{\hat{S}\hat{R}_{y}(-\frac{\pi}{2})\hat{S}^{\dagger},\,\hat{X}_{t},\,\mathbb{I}\}$, with the chirp set by the comb periodicity alone. By introducing and comparing three complementary figures of merit---gate fidelity, state fidelity, and the Knill--Glancy correctability threshold---we have established that a comb with fewer than $\sim 10$ frequency peaks simultaneously prevents high-fidelity gate operation and renders the TF-GKP state uncorrectable. Crucially, we have identified a parameter regime in which the two constraints are compatible: although the gate fidelity is slightly below unity, the residual time-frequency broadening, which acts as the noise in this encoding, remains within the correctable region of the code, so that the TF-Talbot operation constitutes a valid noisy Clifford gate. The visibility of the HOM interference pattern provides a direct and experimentally accessible estimator of the gate fidelity, without requiring full tomography of the phase-matching function; a frequency shift of half the FSR in one arm of the generalised HOM interferometer suffices to distinguish all six logical codewords, and the comb finesse emerges as the central experimental figure of merit. Near-term realisation is within reach using either cavity-enhanced type-II SPDC sources or a broadband source combined with a spatial light modulator, the latter offering a direct route to implementation without the added complexity of a resonant cavity. We recap the simulated performance metrics for performing the TF Talbot operation for different photons platforms producing TF GKP states in Table \ref{tab:performance_summary}.

Several natural extensions of the present work can be identified.
For Clifford operations, the Hadamard gate could be implemented via the fractional Fourier transform, which performs a $\pi/2$ rotation of the chronocyclic Wigner distribution and has already been demonstrated experimentally~\cite{PhysRevLett.130.240801}, though not yet exploited for qubit logical operations on TF-GKP states~\cite{ketterer_quantum_2016}.
Non-Clifford operations for TF-encodings are also within reach: a $T$-gate requires only a relative phase of $\pi/4$, which a single waveshaper can provide without additional complexity.
Beyond single-qubit gates, the TF-Talbot effect could be harnessed to spectrally engineer the correlations of quadrature CV cluster states, following the approach of Ref.~\cite{zhu_hypercubic_2021}.
Rather than through single-photon operations, the TF-Talbot effect can alternatively be realised directly on the collective variables by exploiting the narrow energy conservation of the nonlinear process that generates the photon pair, namely through spatial pump engineering, which directly shapes the joint spectral amplitude of the biphoton state~\cite{douce_direct_2013,zhao_shaping_2015,fabre_generation_2020,francesconi_engineering_2020}. We finally note that the TF-Talbot effect admits a complementary interpretation as a dispersion-correction protocol: a photon that has accumulated a quadratic spectral phase through propagation in a dispersive medium can be restored to its logical codeword, provided the accumulated phase remains below the code noise threshold.

\section*{Acknowledgments}
T. Pousset acknowledges funding from IMT, l’Institut Carnot TSN and the Fondation Mines-Télécom. N. Fabre acknowledges support from the European Union’s Horizon Europe research and innovation programme under the project Quantum Secure Network Partnership (QSNP, grant agreement No~101114043). N. Fabre acknowledges fruitful and long-term discussions with Pérola Milman, Maria Amanti, Sara Ducci, and Giorgio Maltese on the preliminary ideas of this manuscript. L. Labonté would like to thank A. Martin and S. Tanzilli for their valuable insights into the experimental feasibility of the project.

\section*{Data availability}
The data are available from the authors upon reasonable request. The code can be found in~\cite{repo_key}.

\appendix

\section{Inner products of TF-GKP states}\label{app:inner-products}

In Sec.~\ref{sec:physical_gkp}, we introduced physical TF-GKP states parametrised by a finite peak width $\sigma$ and envelope width $\kappa$, and computed the inner products between the logical codewords as functions of these parameters. Here we provide the detailed derivations of the associated normalisation factors and inner products.
\par We start with the normalization factors of the $\ket{0_\omega}$ and $\ket{1_\omega}$ states.
The factor $N_{0_\omega}$ is calculated as follows:

\begin{equation}
    \begin{aligned}
        \frac{1}{N_{0_\omega}^2} &= \int\mathrm{d}\omega e^{-\frac{\omega^2}{\kappa^2}}\abs{\sum_{n\in\mathbb{Z}}e^{-\frac{(\omega-2n\overline\omega)^2}{2\sigma^2}}}^2,\\
        &= \int\mathrm{d}\omega e^{-\frac{\omega^2}{\kappa^2}}\sum_{n,m\in\mathbb{Z}}e^{-\frac{(\omega-2n\overline\omega)^2}{2\sigma^2} -\frac{(\omega-2m\overline\omega)^2}{2\sigma^2}},\\
        &= \sum_{n,m\in\mathbb{Z}}e^{-\frac{(n-m)^2\overline\omega^2}{\sigma^2}}\int\mathrm{d}\omega e^{-\frac{\omega^2}{\kappa^2}}e^{-\frac{(\omega-(n+m)\overline\omega)^2}{\sigma^2}},\\
        &= \sum_{n,m\in\mathbb{Z}}e^{-\frac{(n-m)^2\overline\omega^2}{\sigma^2}}e^{-\frac{(n+m)^2\overline\omega^2}{\kappa^2+\sigma^2}}\int\mathrm{d}\omega e^{-\frac{\kappa^2+\sigma^2}{\kappa^2\sigma^2}(\omega - \frac{\kappa^2(n+m)}{\kappa^2+\sigma^2}\overline\omega)^2},\\
        &= \sqrt{\pi\frac{\kappa^2\sigma^2}{\kappa^2+\sigma^2}}\sum_{n,m\in\mathbb{Z}}e^{-\frac{(n-m)^2\overline\omega^2}{\sigma^2}}e^{-\frac{(n+m)^2\overline\omega^2}{\kappa^2+\sigma^2}}.
    \end{aligned}
\end{equation}

This expression can be further simplified when the peak width is negligible compared to the FSR, $\sigma\ll\overline\omega$, in which case the cross terms in the double sum above can be neglected.
This physically corresponds to accounting only for overlapping peaks when taking the modulus square of the wavefunction, and thus neglecting products of non-overlapping peaks such as the pairs $n$ and $n+1$, $n$ and $n-1$, etc.
The simplification can be written as:
\begin{equation}
    \begin{aligned}
        \frac{1}{N_{0_\omega}^2} &\approx \sqrt{\pi\frac{\kappa^2\sigma^2}{\kappa^2+\sigma^2}}\sum_{n\in\mathbb{Z}}e^{-\frac{4n^2\overline\omega^2}{\kappa^2+\sigma^2}},\\
        &= \sqrt{\pi\frac{\kappa^2\sigma^2}{\kappa^2+\sigma^2}}\theta_3(0|i\frac{4\overline\omega^2}{\pi(\kappa^2+\sigma^2)}),
    \end{aligned}
\end{equation}
where $\theta_3(z|\tau)=\sum_{n\in\mathbb{Z}}e^{i\pi\tau n^2 + i\pi zn}$ is the Jacobi theta function~\cite{national_institute_of_standards_and_technology_nist_2010}.
The above expression can be further simplified when the envelope width is much larger than the FSR, meaning $\kappa\gg\overline\omega$, by approximating the sum with an integral:
\begin{equation}
    \begin{aligned}
        \frac{1}{N_{0_\omega}^2} &\approx \sqrt{\pi\frac{\kappa^2\sigma^2}{\kappa^2+\sigma^2}}\sum_{n\in\mathbb{Z}}e^{-\frac{4n^2\overline\omega^2}{\kappa^2+\sigma^2}},\\
        &\approx \sqrt{\pi\frac{\kappa^2\sigma^2}{\kappa^2+\sigma^2}}\int\mathrm{d}x e^{-\frac{4x^2\overline\omega^2}{\kappa^2+\sigma^2}},\\
        &= \sqrt{\pi\frac{\kappa^2\sigma^2}{\kappa^2+\sigma^2}}\sqrt{\pi\frac{\kappa^2 + \sigma^2}{4\overline\omega^2}},\\
        &= \pi\frac{\kappa\sigma}{2\overline\omega},
    \end{aligned}\label{eq:N_0_omega}
\end{equation}
where the error term can be calculated using the Euler-Maclaurin formula~\cite{national_institute_of_standards_and_technology_nist_2010}.
\par Similarly, the $N_{1_\omega}$ normalization factor can be calculated as follows:

\begin{equation}
    \begin{aligned}
        \frac{1}{N_{1_\omega}^2} &= \int\mathrm{d}\omega e^{-\frac{\omega^2}{\kappa^2}}\abs{\sum_{n\in\mathbb{Z}}e^{-\frac{(\omega-(2n+1)\overline\omega)^2}{2\sigma^2}}}^2,\\
        &= \sqrt{\pi\frac{\kappa^2\sigma^2}{\kappa^2+\sigma^2}}\sum_{n,m\in\mathbb{Z}}e^{-\frac{(n-m)^2\overline\omega^2}{\sigma^2}}e^{-\frac{(n+m+1)^2\overline\omega^2}{\kappa^2+\sigma^2}},\\
        &\approx \sqrt{\pi\frac{\kappa^2\sigma^2}{\kappa^2+\sigma^2}}\sum_{n\in\mathbb{Z}}e^{-\frac{(2n+1)^2\overline\omega^2}{\kappa^2+\sigma^2}},\\
        &\approx \sqrt{\pi\frac{\kappa^2\sigma^2}{\kappa^2+\sigma^2}}\int\mathrm{d}xe^{-(2x+1)^2\frac{\overline\omega^2}{\kappa^2+\sigma^2}},\\
        &= \pi\frac{\kappa\sigma}{2\overline\omega},
    \end{aligned}
\end{equation}

where the second-to-third line uses the narrow-peak approximation $\sigma\ll\overline\omega$, and the third-to-fourth line uses the large-envelope approximation $\kappa\gg\overline\omega$.
Finally, both normalization factors can be written as $N_{0_\omega}\approx N_{1_\omega} \approx \sqrt{2\overline\omega/(\pi\kappa\sigma)}$.
\par The same calculation technique can be applied for $N_{0_t}$ and $N_{1_t}$, which leads to:
\begin{equation}
    \begin{aligned}
        \frac{1}{N_{0_t}^2} &= \int \mathrm{d}t\abs{\sum_{k\in\mathbb{Z}}e^{-2\pi^2 k^2 \frac{\sigma^2}{\overline\omega^2}}e^{-\frac{(t-2k\frac{\pi}{\overline\omega})^2\kappa^2}{2}}}^2,\\
        &=\sqrt{\frac{\pi}{\kappa^2}}\sum_{k,l\in\mathbb{Z}}e^{-2\pi^2(k^2+l^2)\frac{\sigma^2}{\overline\omega^2}}e^{-\kappa^2(k-l)^2\frac{\pi^2}{\overline\omega^2}},\\
        &\approx \sqrt{\frac{\pi}{\kappa^2}}\sum_{k\in\mathbb{Z}}e^{-4\pi^2k^2\frac{\sigma^2}{\overline\omega^2}},\\
        &\approx \sqrt{\frac{\pi}{\kappa^2}}\sqrt{\frac{\overline\omega^2}{4\pi\sigma^2}}=\frac{\overline\omega}{2\kappa\sigma},\label{eq:norm_2_t}
    \end{aligned}
\end{equation}
\begin{equation}
    \begin{aligned}
        \frac{1}{N_{1_t}^2} &= \int\mathrm{d}t\abs{\sum_{k\in\mathbb{Z}}e^{-\frac{(2k+1)^2\pi^2}{2}\frac{\sigma^2}{\overline\omega^2}}e^{-\frac{\kappa^2(t-(2k+1)\frac{\pi}{\overline\omega})^2}{2}}}^2,\\
        &= \sqrt{\frac{\pi}{\kappa^2}}\sum_{k,l\in\mathbb{Z}}e^{-\frac{(2k+1)^2+(2l+1)^2}{2}\frac{\pi^2\sigma^2}{\overline\omega^2}}e^{-\kappa^2(k-l)^2\frac{\pi^2}{\overline{\omega}^2}},\\
        &\approx \sqrt{\frac{\pi}{\kappa^2}}\sum_{k\in\mathbb{Z}}e^{-(2k+1)^2\frac{\pi^2\sigma^2}{\overline\omega^2}},\\
        &\approx \sqrt{\frac{\pi}{\kappa^2}}\sqrt{\frac{\overline\omega^2}{4\pi\sigma^2}} = \frac{\overline\omega}{2\kappa\sigma},\label{eq:norm_1_t}
    \end{aligned}
\end{equation}
where for the third line, we used the large-envelope approximation, and for the fourth line, we used the narrow-peak-width approximation, in both Eq.~(\ref{eq:norm_1_t}) and Eq.~(\ref{eq:norm_2_t}).
Finally, the normalization factors can be written as $N_{0_t}\approx N_{1_t} \approx \sqrt{2\kappa\sigma/\overline\omega}$.

\par We then calculate the overlaps between the $\ket{0_\omega}$ and the $\ket{1_\omega}$ states and between the $\ket{0_t}$ and the $\ket{1_t}$ states.
The first overlap can be written as:
\begin{multline}
    \braket{0_\omega}{1_\omega} \\
    = N_{0_\omega}N_{1_\omega}\int\mathrm{d}\omega e^{-\frac{\omega^2}{\kappa^2}}\sum_{n,m\in\mathbb{Z}}e^{-\frac{(\omega-2n\overline\omega)^2}{2\sigma^2}-\frac{(\omega-(2m+1)\overline\omega)^2}{2\sigma^2}},\\
    = N_{0_\omega}N_{1_\omega}\sum_{n,m\in\mathbb{Z}}e^{-\frac{(n-m-\frac{1}{2})^2\overline\omega^2}{\sigma^2}}\int\mathrm{d}\omega e^{-\frac{\omega^2}{\kappa^2}}e^{-\frac{(\omega-(n+m+\frac{1}{2})\overline\omega)^2}{\sigma^2}},\\
    = N_{0_\omega}N_{1_\omega}\sum_{n,m\in\mathbb{Z}}e^{-\frac{(n-m-\frac{1}{2})^2\overline\omega^2}{\sigma^2}}e^{-\frac{(n+m+\frac{1}{2})^2\overline\omega^2}{\kappa^2+\sigma^2}}\\
    \times\int\mathrm{d}\omega e^{-\frac{\kappa^2+\sigma^2}{\kappa^2\sigma^2}(\omega-\frac{\kappa^2}{\kappa^2+\sigma^2}(n+m+\frac{1}{2})\overline\omega)^2},\\
    = N_{0_\omega}N_{1_\omega}\sqrt{\pi\frac{\kappa^2\sigma^2}{\kappa^2+\sigma^2}}\sum_{n,m\in\mathbb{Z}}e^{-\frac{(n-m-\frac{1}{2})^2\overline\omega^2}{\sigma^2}}e^{-\frac{(n+m+\frac{1}{2})^2\overline\omega^2}{\kappa^2+\sigma^2}}\\
    \approx N_{0_\omega}N_{1_\omega}\sqrt{\pi\frac{\kappa^2\sigma^2}{\kappa^2+\sigma^2}}e^{-\frac{\overline\omega^2}{4\sigma^2}}\sum_{n\in\mathbb{Z}}e^{-\frac{(2n+\frac{1}{2})^2\overline\omega^2}{\kappa^2+\sigma^2}}.
\end{multline}
The above expression can be further simplified by assuming that the peaks are narrow, meaning $\sigma\ll\overline\omega$, which leads to neglecting the cross terms in the double sum:
\begin{equation}
    \braket{0_\omega}{1_\omega}\approx N_{0_\omega}N_{1_\omega}\sqrt{\pi\frac{\kappa^2\sigma^2}{\kappa^2+\sigma^2}}e^{-\frac{\overline\omega^2}{4\sigma^2}}\sum_{n\in\mathbb{Z}}e^{-\frac{(2n+\frac{1}{2})^2\overline\omega^2}{\kappa^2+\sigma^2}}.\\
\end{equation}
Next, the expression can be simplified by assuming that the envelope is large, meaning that $\kappa\gg\overline\omega$.
Then the sum can be approximated with an integral:
\begin{equation}
    \begin{aligned}
        \braket{0_\omega}{1_\omega} &\approx N_{0_\omega}N_{1_\omega}\sqrt{\pi\frac{\kappa^2\sigma^2}{\kappa^2+\sigma^2}}e^{-\frac{\overline\omega^2}{4\sigma^2}}\int\mathrm{d}x e^{-(2x+\frac{1}{2})^2\frac{\overline\omega^2}{\kappa^2+\sigma^2}},\\
        &= N_{0_\omega}N_{1_\omega}\sqrt{\pi\frac{\kappa^2\sigma^2}{\kappa^2+\sigma^2}}\sqrt{\pi\frac{\kappa^2+\sigma^2}{4\overline\omega^2}}e^{-\frac{\overline\omega^2}{4\sigma^2}},\\
        &=e^{-\frac{\overline\omega^2}{4\sigma^2}},\\
    \end{aligned}
\end{equation}
where the error term of the approximation can be calculated using the Euler-Maclaurin formula~\cite{national_institute_of_standards_and_technology_nist_2010}.
\par The overlap for the temporal states can be calculated in the same manner:
\begin{multline}
    \braket{0_t}{1_t} \\
    = N_{0_t}N_{1_t}\sum_{k,l\in\mathbb{Z}}e^{-\frac{4k^2+(2l+1)^2}{2}\frac{\pi^2\sigma^2}{\overline\omega^2}}\int\mathrm{d}te^{-\frac{(t-2k\frac{\pi}{\overline\omega})^2\kappa^2}{2}-\frac{(t-(2l+1)\frac{\pi}{\overline\omega})^2\kappa^2}{2}},\\
    = N_{0_t}N_{1_t}\sqrt{\frac{\pi}{\kappa^2}}\sum_{k,l\in\mathbb{Z}}e^{-\frac{4k^2+(2l+1)^2}{2}\frac{\pi^2\sigma^2}{\overline\omega^2}}e^{-\frac{(k-l-\frac{1}{2})^2\pi^2\kappa^2}{\overline\omega^2}},\\
    \approx N_{0_t}N_{1_t}\sqrt{\frac{\pi}{\kappa^2}}e^{-\frac{\pi^2\kappa^2}{4\overline\omega^2}}\sum_{k\in\mathbb{Z}}e^{-\frac{4k^2+(2k+1)^2}{2}\frac{\pi^2\sigma^2}{\overline\omega^2}},\\
    = N_{0_t}N_{1_t}\sqrt{\frac{\pi}{\kappa^2}}e^{-\frac{\pi^2\kappa^2}{4\overline\omega^2}}e^{-\frac{\pi^2\sigma^2}{4\overline\omega^2}}\sum_{k\in\mathbb{Z}}e^{-(k+\frac{1}{8})^2\frac{4\pi^2\sigma^2}{\overline\omega^2}},\\
    \approx N_{0_t}N_{1_t}\sqrt{\frac{\pi}{\kappa^2}}\sqrt{\frac{\overline\omega^2}{4\pi\sigma^2}}e^{-\frac{\pi^2\kappa^2}{4\overline\omega^2}} = e^{-\frac{\pi^2\kappa^2}{4\overline\omega^2}},
\end{multline}
where the second-to-third equality uses the large-envelope approximation, and the fourth-to-fifth equality approximates the sum by a Gaussian integral in the limit of small peak width $\sigma$.

\section{Half-Talbot length}\label{appendix: halftalbot}

We derive below the action of the TF-Talbot effect at $\beta = \beta_T/2$ and $\beta = 3\beta_T/2$ on the ideal TF-GKP codewords.

\paragraph{Action on $\ket{\overline{0}_t}$.}
The temporal wavefunction of the $\ket{\overline{0}_t}$ state at $\beta = 0$ reads
\begin{equation}
    \tilde\psi_{\ket{\overline0_t}}(t,0) = \braket{t}{\overline{+}_\omega} = \sum_{k\in\mathbb{Z}} \delta\!\left(t - \frac{2k\pi}{\overline\omega}\right) = \sum_{n\in\mathbb{Z}} e^{in\overline\omega t}.
\end{equation}
After propagation through a dispersive element of chirp $\beta = \beta_T/2$, the wavefunction becomes
\begin{equation}
    \tilde\psi_{\ket{\overline0_t}}\!\left(t,\tfrac{\beta_T}{2}\right) = \sum_{n\in\mathbb{Z}} e^{in\overline\omega t}\,e^{i\frac{\pi}{2}n^2} = \sum_{n\in\mathbb{Z}} i^{n^2}\,e^{in\overline\omega t},
\end{equation}
or equivalently in the frequency domain:
\begin{equation}
    \psi_{\ket{\overline0_t}}\!\left(\omega,\tfrac{\beta_T}{2}\right) = \sum_{n\in\mathbb{Z}} i^{n^2}\,\delta(\omega - n\overline\omega).
\end{equation}
The phase factor $i^{n^2}$ takes two distinct values depending on the parity of $n$. For even $n = 2p$, one has $n^2 = 4p^2$ and thus $i^{n^2} = 1$. For odd $n = 2p+1$, one has $n^2 = 4p^2 + 4p + 1$ and thus $i^{n^2} = i$. Separating the sum into even and odd contributions gives
\begin{equation}
    \psi_{\ket{\overline0_t}}\!\left(\omega,\tfrac{\beta_T}{2}\right) = \sum_{p\in\mathbb{Z}} \delta(\omega - 2p\overline\omega) + i\sum_{p\in\mathbb{Z}} \delta\!\left(\omega - (2p+1)\overline\omega\right),
\end{equation}
in which we recognise the $\ket{\overline{0}_\omega}$ and $\ket{\overline{1}_\omega}$ components, so that
\begin{equation}
    \begin{aligned}
        \psi_{\ket{\overline0_t}}\!\left(\omega,\tfrac{\beta_T}{2}\right)
        &= \psi_{\ket{\overline0_\omega}}(\omega,0) + i\,\psi_{\ket{\overline1_\omega}}(\omega,0)\\
        &= \sqrt{2}\,\psi_{\ket{\overline{+i}_\omega}}(\omega,0).
    \end{aligned}
\end{equation}
Again, the factor $1/\sqrt{2}$ is a convention imposed at the qubit level only when one restricts to the finite-dimensional logical subspace spanned by normalizable TF-GKP states. Expressing the same result in the time domain yields
\begin{equation}
    \begin{aligned}
        \tilde\psi_{\ket{\overline0_t}}\!\left(t,\tfrac{\beta_T}{2}\right)
        &= \tilde\psi_{\ket{\overline+_t}}(t,0) + i\,\tilde\psi_{\ket{\overline-_t}}(t,0)\\
        &= (1+i)\,\tilde\psi_{\ket{\overline0_t}}(t,0) + (1-i)\,\tilde\psi_{\ket{\overline1_t}}(t,0)\\
        &= (1+i)\,\tilde\psi_{\ket{\overline{-i}_t}}(t,0),
    \end{aligned}
\end{equation}
recovering the result of Eq.~(\ref{equation:halftalbot}). We note that the global phase factor $(1+i) \propto e^{i\pi/4}$ is state-dependent: it differs when the input is $\ket{\overline{1}_t}$, as shown below.

\paragraph{Action on $\ket{\overline{1}_t}$.}
Applying the same procedure to the $\ket{\overline{1}_t}$ state, whose frequency wavefunction carries an additional $(-1)^n$ factor, gives
\begin{equation}
    \psi_{\ket{\overline1_t}}\!\left(\omega,\tfrac{\beta_T}{2}\right) = \sum_{n\in\mathbb{Z}} (-1)^n\,i^{n^2}\,\delta(\omega - n\overline\omega).
\end{equation}
For even $n$ the phase is $(-1)^n i^{n^2} = 1$, and for odd $n$ it is $(-1)^n i^{n^2} = -i$, so that
\begin{equation}
    \begin{aligned}
        \psi_{\ket{\overline1_t}}\!\left(\omega,\tfrac{\beta_T}{2}\right)
        &= \psi_{\ket{\overline0_\omega}}(\omega,0) - i\,\psi_{\ket{\overline1_\omega}}(\omega,0)\\
        &= \sqrt{2}\,\psi_{\ket{\overline{-i}_\omega}}(\omega,0).
    \end{aligned}
\end{equation}
In the time domain, this reads
\begin{equation}
    \begin{aligned}
        \tilde\psi_{\ket{\overline1_t}}\!\left(t,\tfrac{\beta_T}{2}\right)
        &= \tilde\psi_{\ket{\overline+_t}}(t,0) - i\,\tilde\psi_{\ket{\overline-_t}}(t,0)\\
        &= (1-i)\,\tilde\psi_{\ket{\overline0_t}}(t,0) + (1+i)\,\tilde\psi_{\ket{\overline1_t}}(t,0)\\
        &= (1-i)\,\tilde\psi_{\ket{\overline{+i}_t}}(t,0),
    \end{aligned}
\end{equation}
recovering Eq.~(\ref{equation:halftalbot}). We therefore confirm that the half-Talbot evolution implements the $\hat{S}\hat{R}_{y}(-\frac{\pi}{2})\hat{S}^{\dagger}$ gate up to a state-dependent phase.

\section{Knill--Glancy criterion for TF-GKP states}\label{app:knill-glancy}

In Sec.~\ref{sec:knill-glancy}, we calculate the probability of error of a physical TF-GKP state's stabilization using a Steane circuit as in~\cite{glancy_error_2006}.
For GKP states encoded in quadratures and stabilization with a Steane circuit, the probability of having a logical error after a stabilization cycle is the probability of measuring a displacement of more than $\sqrt{\pi}/6$, corresponding to a sixth of the displacement necessary to perform an $\hat{X}$ gate or a $\hat{Z}$ gate.
For TF-GKP states, the displacement necessary for an $\hat{X}_\omega=\hat{Z}_t$ gate is a frequency shift of $\overline\omega$, and the displacement necessary for a $\hat{Z}_\omega=\hat{X}_t$ gate is a time shift of $\pi/\overline\omega$.
As in~\cite{glancy_error_2006}, we calculate the portion of the wavefunction contained in the $[-\overline\omega/6,\overline\omega/6[\times[-\pi/(6\overline\omega),\pi/(6\overline\omega)[$ region in the modular phase space.
\par We make use of the formalism of modular variables~\cite{ketterer_quantum_2016} to calculate the wavefunction of a realistic TF-GKP state in the modular phase space.
The basis associated with modular frequency and time is written:

\begin{equation}
    \ket{\xi,\tilde{t}} = \sqrt{\frac{\overline\omega}{\pi}}\sum_{n\in\mathbb{Z}}e^{2i\tilde{t}n\overline\omega}\ket{\xi+2n\overline\omega},
\end{equation}

where $\xi\in[-\overline\omega/2,3\overline\omega/2]$ is the modular frequency and $\tilde{t}\in[-\pi/(2\overline\omega),\pi/(2\overline\omega)]$ is the modular time.
In this basis, the $\ket{0_\omega}$ state can be written as:
\begin{equation}
    \begin{aligned}
        \ket{0_\omega} &= \int_{-\frac{\overline\omega}{2}}^{\frac{3\overline\omega}{2}}\mathrm{d}\xi\int_{-\frac{\pi}{2\overline\omega}}^{\frac{\pi}{2\overline\omega}}\mathrm{d}\tau \overline{\psi}_{0_\omega}(\xi,\tau)\ket{\xi,\tau},
    \end{aligned}
\end{equation}
where $\overline\psi(\xi,\tau) = \sqrt{\frac{\overline\omega}{\pi}}\sum_{n\in\mathbb{Z}}\psi_{0_\omega}(\xi+2n\overline\omega)e^{-2in\tau\overline\omega}$ is the Zak transform of $\psi_{0_\omega}(\omega) = \braket{\omega}{0_\omega}$.
Using Eq.~(\ref{eq:physical_0_1}), $\overline\psi(\xi,\tau)$ can be written as:
\begin{equation}
    \begin{aligned}
        \overline\psi_{0_\omega}(\xi,\tau) &\propto\sum_{(n,k)\in\mathbb{Z}^{2}}e^{-2in\tau\overline\omega}e^{-\frac{(\xi+2n\overline\omega)^2}{2\kappa^2}}e^{-\frac{(\xi-2(k-n)\overline\omega)^2}{2\sigma^2}}\\
        &\propto \sum_{(n,l)\in\mathbb{Z}^{2}}e^{-2in\tau\overline\omega}e^{-\frac{(\xi+2n\overline\omega)^2}{2\kappa^2}}e^{-\frac{(\xi-2l\overline\omega)^2}{2\sigma^2}}\\
        &\propto\sum_{(m,l)\in\mathbb{Z}^{2}}e^{-\frac{(\xi-2l\overline\omega)^2}{2\sigma^2}}e^{-\frac{\kappa^2(\tau-m\frac{\pi}{\overline\omega})^2}{2}}e^{i\xi(\tau-m\frac{\pi}{\overline\omega})},
    \end{aligned}
\end{equation}
where the last line follows from the Poisson summation formula~\cite{national_institute_of_standards_and_technology_nist_2010}, and where $l = k - n \in \mathbb{Z}$ in the second line. To simplify further, we assume that the envelope $\kappa$ is large compared to the FSR $\overline{\omega}$, so that the phase factor $e^{i\xi(\tau-m\pi/\overline\omega)}$ reduces to unity at leading order. The Zak-basis wavefunction then takes the form:
\begin{equation}
    \overline\psi_{0_\omega}(\xi,\tau) = \overline{N}_{0_\omega}\sum_{(m,l)\in\mathbb{Z}^{2}}e^{-\frac{(\xi-2l\overline\omega)^2}{2\sigma^2}}e^{-\frac{\kappa^2(\tau-m\frac{\pi}{\overline\omega})^2}{2}},
\end{equation}
where $\overline{N}_{0_\omega}$ is a normalization factor that can be written as $\overline N_{0_\omega} = \sqrt{\frac{\kappa}{\pi\sigma}}$.
\par Using the results of~\cite{ketterer_quantum_2016}, we can write states $\ket{\phi}$ in the physical TF-GKP codespace, associated with the ideal codestate $\alpha\ket{\overline 0_{\omega}} + \beta\ket{\overline 1_{\omega}}$, as:
\begin{equation}
    \ket{\phi} = \int_{-\frac{\overline\omega}{2}}^{\frac{\overline\omega}{2}}\mathrm{d}\xi\int_{-\frac{\pi}{2\overline\omega}}^{\frac{\pi}{2\overline\omega}}\mathrm{d}\tau f(\xi,\tau)(\alpha\ket{\xi,\tau} + \beta\ket{\xi+\overline\omega,\tau}),
\end{equation}
where $f(\xi,\tau)$ encodes the envelope and the peak width.
The probability of measuring the state at position ($\xi$, $\tau$) in modular phase space is given by:
\begin{equation}
    P(\xi,\tau) = \abs{f(\xi,\tau)}^2.
\end{equation}
We emphasize that measuring $\xi$ and $\tau$ does not violate the Heisenberg uncertainty principle, as only the modular time and modular frequency are measured.
\par We can write the probability in terms of the Zak transform as:
\begin{multline}
    \abs{f(\xi,\tau)}^2 = \abs{\overline\psi_{0_\omega}(\xi,\tau)}^2+\abs{\overline\psi_{0_\omega}(\xi+\overline\omega,\tau)}^2,\\
    =\overline{N}_{0_\omega}^2\abs{\sum_{m\in\mathbb{Z}}e^{-\frac{\kappa^2(\tau-m\frac{\pi}{\overline\omega})^2}{2}}}^2\\
    \times\left(\abs{\sum_{l\in\mathbb{Z}}e^{-\frac{(\xi-2l\overline\omega)^2}{2\sigma^2}}}^2+\abs{\sum_{l\in\mathbb{Z}}e^{-\frac{(\xi-(2l+1)\overline\omega)^2}{2\sigma^2}}}^2\right).\label{eq:f_xi_tau}
\end{multline}
The first sum can be written as:
\begin{equation}
    \begin{aligned}
        \abs{\sum_{m\in\mathbb{Z}}e^{-\frac{\kappa^2(\tau-m\frac{\pi}{\overline\omega})^2}{2}}}^2 &= \sum_{(n,m)\in\mathbb{Z}^{2}}e^{-\frac{\kappa^2(\tau-m\frac{\pi}{\overline\omega})^2}{2}}e^{-\frac{\kappa^2(\tau-n\frac{\pi}{\overline\omega})^2}{2}} ,\\
        &= \sum_{(n,m)\in\mathbb{Z}^{2}}e^{-\frac{\kappa^2}{4}(m-n)^2\frac{\pi^2}{\overline\omega^2}}e^{-\kappa^2(\tau-(n+m)\frac{\pi}{2\overline\omega})^2}.\\
    \end{aligned}
\end{equation}
The last factor of Eq.~(\ref{eq:f_xi_tau}) can be simplified into:
\begin{multline}
    \left(\abs{\sum_{l\in\mathbb{Z}}e^{-\frac{(\xi-2l\overline\omega)^2}{2\sigma^2}}}^2+\abs{\sum_{l\in\mathbb{Z}}e^{-\frac{(\xi-(2l+1)\overline\omega)^2}{2\sigma^2}}}^2\right)\\
    =\sum_{(l,k)\in\mathbb{Z}^{2}}e^{-\frac{(k-l)^2\overline\omega^2}{\sigma^2}}\left(e^{-\frac{(\xi-(k+l)\overline\omega)^2}{\sigma^2}}+e^{-\frac{(\xi-(k+l+1)\overline\omega)^2}{\sigma^2}}\right),\\
    =\sum_{(u,v)\in\mathbb{Z}^{2}}e^{-\frac{u^2\overline\omega^2}{\sigma^2}}\left(e^{-\frac{(\xi-(u+2v)\overline\omega)^2}{\sigma^2}}+e^{-\frac{(\xi-(u+2v+1)\overline\omega)^2}{\sigma^2}}\right),\\
    =\sum_{(u,v)\in\mathbb{Z}^{2}}e^{-\frac{u^2\overline\omega^2}{\sigma^2}}e^{-\frac{(\xi-(u+v)\overline\omega)^2}{\sigma^2}} = \sum_{(k,l)\in\mathbb{Z}^{2}}e^{-\frac{(k-l)^2\overline\omega^2}{\sigma^2}}e^{-\frac{(\xi-l\overline\omega)^2}{\sigma^2}}.
\end{multline}
Finally, the probability of measuring $(\xi,\tau)$ for the modular frequency and time of the TF-GKP physical qubit is:

\begin{multline}
   P(\xi,\tau) = \overline N_{0_\omega}^2\\
  \sum_{n,m,k,l\in\mathbb{Z}}e^{-\frac{\kappa^2}{4}(m-n)^2\frac{\pi^2}{\overline\omega^2}}e^{-\frac{(k-l)^2\overline\omega^2}{\sigma^2}}e^{-\kappa^2(\tau-(n+m)\frac{\pi}{2\overline\omega})^2}e^{-\frac{(\xi-l\overline\omega)^2}{\sigma^2}}.
\end{multline}

Using the results of Ref.~\cite{glancy_error_2006}, we calculate the probability of error after a stabilization cycle of a TF-GKP state using a Steane-type correction circuit with a noisy ancilla.
The probability of having no error can be written as:
\begin{multline}
    P_{\text{no error}}=\int_{-\frac{\overline\omega}{6}}^{\frac{\overline\omega}{6}}\mathrm{d}\xi\int_{-\frac{\pi}{6\overline\omega}}^{\frac{\pi}{6\overline\omega}}\mathrm{d}\tau P(\xi,\tau),\\
    =\overline N_{0_\omega}^2\sum_{n,m,k,l\in\mathbb{Z}}e^{-\frac{\kappa^2}{4}(m-n)^2\frac{\pi^2}{\overline\omega^2}}e^{-\frac{(k-l)^2\overline\omega^2}{\sigma^2}}\\
    \times\sqrt{\frac{\pi}{4\kappa^2}}(\erf((3(m+n)+1)\frac{\pi\kappa}{6\overline\omega})-\erf((3(m+n)-1)\frac{\pi\kappa}{6\overline\omega}))\\
    \times\sqrt{\frac{\pi\sigma^2}{4}}(\erf((l+\frac{1}{6})\frac{\overline\omega}{\sigma})-\erf((l-\frac{1}{6})\frac{\overline\omega}{\sigma})).
\end{multline}
The expression can be simplified in the approximation of narrow peak width $\sigma\ll\overline\omega$ and large envelope $\kappa\gg\overline\omega$, and then corresponds to Eq.~(\ref{probaerror}).

\section{Calculation of the chronocyclic Wigner distribution for ideal TF-GKP states}\label{app:wigner_ideal}

We detail here the calculation of the chronocyclic Wigner distribution for ideal TF-GKP states, and derive how a dispersive element transforms it, providing the phase-space picture underlying the HOM signatures discussed in Sec.~\ref{sec:HOM-theory}.

\paragraph{Wigner distribution of the logical codewords.}
For a pure state with frequency wavefunction $\psi(\omega)$, the chronocyclic Wigner distribution is defined as

\begin{equation}
    W_{\ket{\psi}}(\mu,\tau) = \int_{-\infty}^{\infty} d\omega\, e^{2i\omega\tau}\,\psi(\mu-\omega)\,\psi^*(\omega+\mu),
\end{equation}

where $\tau$ and $\mu$ are the time and frequency shifts of the generalised HOM protocol.
Starting from the $\ket{\overline{0}_\omega}$ state, the Wigner distribution reads
\begin{equation}
    W_{\ket{\overline{0}_\omega}}(\mu,\tau) = \sum_{(s,t)\in\mathbb{Z}^{2}} \delta\!\left(\mu - \overline{\omega}(t+s)\right) e^{2i\overline{\omega}\tau(t-s)}.
\end{equation}
Performing the change of variables $k = t+s$ and recognising the Poisson summation formula
$\sum_{s\in\mathbb{Z}} e^{-4i\overline{\omega}\tau s} = \frac{\pi}{2\overline{\omega}}\sum_{s\in\mathbb{Z}}\delta\!\left(\tau - s\frac{\pi}{2\overline{\omega}}\right)$,
one obtains

\begin{equation}
    W_{\ket{\overline{0}_\omega}}(\mu,\tau) = \sum_{(s,k)\in\mathbb{Z}^{2}} \delta\!\left(\tau - s\frac{\pi}{2\overline{\omega}}\right) \delta(\mu - \overline{\omega}k)\, e^{i\pi sk}.
\end{equation}

The sign of each peak is determined by the parity of $s$ and $k$:
\begin{equation}
    e^{i\pi sk} = \begin{cases} +1 & \text{if } s \text{ or } k \text{ is even,} \\ -1 & \text{if both } s \text{ and } k \text{ are odd.} \end{cases}
\end{equation}
For the $\ket{\overline{1}_\omega}$ state, the chronocyclic Wigner distribution acquires an additional factor $e^{i\pi s}$,

\begin{equation}
    W_{\ket{\overline{1}_\omega}}(\mu,\tau) = \sum_{(s,k)\in\mathbb{Z}^{2}} \delta\!\left(\tau - s\frac{\pi}{2\overline{\omega}}\right) \delta(\mu - \overline{\omega}k)\, e^{i\pi s(k+1)},
\end{equation}

which is equivalent to a unit shift along the $k$ direction in the ideal comb.

\paragraph{Effect of dispersion.}
Propagation through a dispersive medium of parameter $\beta$ multiplies the frequency wavefunction by $e^{i\beta\omega^2}$, so that $\psi(\omega)\big|_\beta = e^{i\beta\omega^2}\psi(\omega)\big|_{\beta=0}$. Substituting into the Wigner distribution and simplifying the resulting phase,

\begin{align}
    W_{\ket{\psi}}(\tau,\mu,\beta) &=
     \int_{-\infty}^{\infty} d\omega\, e^{2i\omega(\tau - 2\beta\mu)}\,\psi(\omega{-}\mu,0)\,\psi^*(\omega{+}\mu,0) \notag\\
    &= W(\tau - 2\beta\mu,\,\mu,\,0),
\end{align}

showing that dispersion acts as a $\mu$-dependent shear of the Wigner distribution along the $\tau$ axis. Writing $\beta = b\,\beta_T$ with $\beta_T = \pi/\overline{\omega}^2$, the sheared Wigner distribution of the $\ket{\overline{+}_\omega}$ state becomes

\begin{equation}
    W_{\ket{\overline{+}_\omega}}(\tau,\mu)\big|_\beta = \frac{1}{4}\sum_{(s,k)\in\mathbb{Z}^{2}} \delta\!\left(\tau - (kb+s)\frac{\pi}{\overline{\omega}}\right) \delta\!\left(\mu - \frac{\overline{\omega}}{2}k\right) e^{i\pi sk}.
\end{equation}

At $b=1$ (\textit{i.e}.\ $\beta = \beta_T$), the peak originally located at $(s,k)$ is displaced to the position $s + k$, implementing the $\hat{X}_t$ logical gate; at $b = 1/2$ (\textit{i.e}.\ $\beta = \beta_T/2$), the displacement is $s + k/2$, implementing the $\hat{S}\hat{R}_{y}(-\frac{\pi}{2})\hat{S}^{\dagger}$ gate.

\bibliography{name1}

\end{document}